\documentclass[rppweb]{pdg}

\usepackage[]{makeidx}

\makeindex

\def\mstopa{M_{\widetilde t_1}}
\def\mstopb{M_{\widetilde t_2}}
\def\msusyy{M_{\rm S}^2}
\AtBeginDocument{%
    \renewcommand*{\Ref}[1]{%
       Ref.~\cite{#1}%
    }%
}

\begin{document}

\begin{center}

\Large\bf\boldmath
Supersymmetry, Part I (Theory)

\unboldmath
\end{center}
\vspace{0.4cm}
\begin{center}

Ben~Allanach\footnote{Electronic address: b.c.allanach@damtp.cam.ac.uk}\\
 Department of Applied Mathematics and Theoretical Physics, \\University of
 Cambridge,  
 Wilberforce Road, Cambridge, CB3 0WA, United Kingdom\\[0.2cm]

and \\[0.2cm]

Howard E.~Haber\footnote{Electronic address: haber@scipp.ucsc.edu} \\
Santa Cruz Institute for Particle Physics, University of California, \\
1156 High Street, Santa Cruz, CA 95064 USA
\end{center}

\vspace{0.3cm}

\centerline{\bf \large Abstract} 
\vspace{0.2cm}
This is a review of the theoretical aspects of the supersymmetric extension of the Standard Model of
particle physics, extracted from
Chapter 87 of the \textit{2026 Review of Particle Physics},
F.~Takahashi et al. (Particle Data Group), Int. J. Mod. Phys. {\bf A41},
2630011 (2026).
 The companion review, co-authored by
 M. D’Onofrio and F. Moortgat, 
``Supersymmetry, Part II (Experiment),'' can be found in Chapter 88 of the 
\textit{2026 Review of Particle Physics} (\textit{op.\!~cit.}).

\vspace{0.8cm}

\centerline{\bf \large Table of Contents} 
\vspace{0.2cm}

\tableofcontents
\medskip

\section{Introduction}
\label{susy1:sec:Intro}
\medskip

\index{SUSY, supersymmetry}%
\index{supersymmetry, SUSY}%
Supersymmetry (SUSY) is a generalization of the space-time symmetries of
quantum field theory that transforms fermions into bosons and
vice versa\cite{susy1:KaneShifman}.
The existence of such a non-trivial extension of the Poincar\'e
symmetry of ordinary quantum field theory was initially
surprising, and its form is highly constrained by theoretical
principles\cite{Haag:1974qh,*Coleman:1967ad}.  For example, supersymmetry has been used as a theoretical laboratory for studying nonperturbative aspects of strongly-coupled quantum field theories (e.g., see Refs.~\cite{susy1:terning,susy1:bertolini}).

SUSY also provides a framework for the unification
of particle physics and gravity\cite{Nilles:1983ge,susy1:Weinberg00,susy1:Nath}
at the Planck energy scale, 
$M_{\rm P}\sim 10^{19}$~GeV, where
the gravitational interactions become comparable in strength
to the gauge interactions.  Moreover,
super\-symmetry can
{stabilize the hierarchy between}
the energy scale that characterizes
electroweak symmetry breaking, $M_{\rm EW}\sim 100$~GeV,
and the Planck
scale\cite{Witten:1981nf,Dimopoulos:1981zb,*Sakai:1981gr,Susskind:1982mw} {against large
radiative corrections}.
The stability of this large gauge hierarchy
\index{gauge hierarchy}%
with respect to radiative
quantum corrections is not possible to maintain in the
Standard Model (SM) without an unnatural fine-tuning of the
parameters of the fundamental theory at the Planck scale.
In contrast, in a supersymmetric extension of the SM,
it is possible to maintain the gauge hierarchy while
providing a natural framework
for elementary scalar fields.

If super\-symmetry were an exact symmetry of nature, then particles
and their superpartners,
\index{superpartners}%
which differ in spin by half a unit, would be
degenerate in mass.  Since superpartners have not (yet) been observed,
super\-symmetry must be a broken symmetry.  Nevertheless, the
stability of the gauge hierarchy can still be maintained if the
SUSY breaking is soft\cite{Girardello:1981wz,Hall:1990ac,*Jack:1999ud},
and the corresponding SUSY-breaking mass parameters are no
larger than a few TeV.  Whether this is still plausible in light of
recent SUSY searches at the LHC (see Ref.~\cite{SUSYpdgEXP})
will be discussed in \Sec{susy1:sec:exp-MSSM}.

\index{soft supersymmetry breaking}%
In particular, soft-SUSY-breaking
terms of the Lagrangian involve combinations of fields with total
mass dimension of three or less,
with some restrictions on the dimension-three terms as
elucidated in \Ref{Girardello:1981wz}.  The
impact of the soft terms becomes negligible at energy scales much
larger than the size of the SUSY-breaking masses.  Thus,
a theory of weak-scale super\-symmetry, where
the effective scale of super\-symmetry breaking is tied to the scale
of electroweak symmetry breaking, provides a natural framework
for the origin and stability of the gauge
hierarchy\cite{Witten:1981nf,Dimopoulos:1981zb,*Sakai:1981gr,Susskind:1982mw}.

At present, there is no unambiguous experimental evidence for
the breakdown of the SM
at or below the TeV scale.   The expectations for new TeV-scale physics
beyond the SM are based primarily on
three theoretical arguments.  First, in a theory with an elementary scalar
field of mass $m$ and interaction strength $\lambda$
({\it e.g.}, a quartic scalar self-coupling, the square of a gauge coupling or
the square of a Yukawa coupling), the stability with respect
to quantum corrections requires the existence of an energy cutoff roughly of order
$(16\pi^2/\lambda)^{1/2}m$, beyond which new physics must enter\cite{Weisskopf:1939zz}.
A significantly larger energy cutoff would require an unnatural fine-tuning
of parameters that govern the effective low-energy theory.  Applying this
argument to the SM leads to an expectation of new physics
at the TeV scale\cite{Susskind:1982mw}.

Second, the unification of the
three SM gauge
couplings at a very high energy close to the Planck scale
is possible if new physics
beyond the SM (which modifies the running of the
gauge couplings above the electroweak scale) is present.
The minimal supersymmetric extension of
the SM, where superpartner masses lie below a few
TeV, provides an example of successful
gauge coupling unification\cite{Einhorn:1981sx,*Marciano:1981un,susy1:Polonsky}.

Third, the existence of dark matter, which makes up approximately
one quarter of the energy density of the universe, cannot be explained
within the SM of particle physics\cite{Bertone:2004pz}{}.
Remarkably, a stable weakly-interacting massive particle (WIMP)
\index{WIMP, weakly-interacting massive particle}%
\index{weakly-interacting massive particle, WIMP}%
whose mass and interaction rate
are governed by new physics associated with the TeV-scale
can be consistent with the observed density of
dark matter (this is the so-called WIMP miracle, which
is reviewed in \Ref{susy1:miracle}).
\index{WIMP!miracle}%
\index{LSP, lightest supersymmetric particle}%
\index{lightest supersymmetric particle, LSP}%
The lightest supersymmetric particle (LSP), if stable,
is a promising (although not the unique) candidate for
dark matter\cite{Pagels:1981ke,Goldberg:1983nd,Ellis:1983ew,susy1:griest,Steffen:2008qp}{}.  
Further aspects of dark matter can be found in Ref.~\cite{PDGdark}.

\section{Structure of the MSSM} 
\label{susy1:sec:StructMSSM}

\index{MSSM, minimal supersymmetric SM}%
\index{minimal supersymmetric SM, MSSM}%
The minimal supersymmetric extension of the SM (MSSM)
consists of the fields of the
two-Higgs-doublet extension of the SM
and the corresponding superpartner fields\cite{Haber:1984rc,Martin:1997ns,susy1:susybooks,susy1:BaerTata,susy1:DHM}{}.
A particle and its
superpartner together form a supermultiplet.
\index{supermultiplet}%
The corresponding field content of the supermultiplets of the
MSSM and their gauge quantum
numbers are shown in Table~\ref{susy1:tab:MSSM}.  The electric charge
$Q=T_3+\frac{1}{2} Y$ is determined in terms
of the third component of
the weak isospin ($T_3$) and the U(1) weak hypercharge~($Y$).
{\renewcommand{\arraystretch}{1.2}
\begin{pdgwidetable}{|c | c | c | c | c | c | c|}
{The fields of the MSSM and their representations under the 
SU(3)$\times$SU(2)$\times$U(1) gauge group are listed.
For simplicity, only one generation of quarks and leptons is exhibited.  For a given fermion $f$,
we have defined $f_{L,R}\equiv P_{L,R} f$, where $P_{L,R}= \tfrac12(1\mp\gamma_5)f$, and the corresponding left-handed charge-conjugated fermion is given by $f_L^c\equiv P_L f^c =[f_R]^c$.
For each lepton, quark, and Higgs supermultiplet (each denoted by a  hatted upper-case letter),
there is a corresponding antiparticle multiplet of charge-conjugated
fermions and their associated scalar partners\protect\cite{susy1:Lang}. 
\index{gluino}%
\index{gaugino}%
\index{higgsino}%
}{susy1:tab:MSSM}{!ht}
\multicolumn{7}{|c|}{Field Content of the MSSM}\\ \hline
Super- & Super- &  Bosonic & Fermionic & SU(3) & SU(2) & U(1) \\
multiplets & field & fields & partners & & & \\ \hline
gluon/gluino & $\hat V_8$ &  $g$ &  $\widetilde g$ & 8 & 1 &   \hphantom{$-$}0 \\
\hline
gauge boson/ & $\hat V$ & $W^\pm\,,\,W^0$ &  $\widetilde W^\pm\,,\widetilde W^0$ & 1 & 3 & \hphantom{$-$}0 \\
gaugino & $\hat V^\prime$ & $B$ & $\widetilde B$ & 1 & 1 & \hphantom{$-$}0 \\ \hline
slepton/ & $\hat L$ & $(\widetilde\nu_L, \widetilde e^-_L)$ \vphantom{$L_L^{L^L}$} & $(\nu_L,e_L^-)$ & 1 & 2 & $-$1 \\
lepton   & $\hat E^c$ & $\tilde e^+_R$ & $e_L^c$  & 1 & 1 & \hphantom{$-$}2 \\ \hline
squark/ & $\hat Q$ & $(\widetilde u_L,\widetilde d_L)$ \vphantom{$L_L^{L^L}$}  & $(u_L,d_L)$ & 3 & 2 & \hphantom{$-$}1/3 \\
quark   & $\hat U^c$ & $\widetilde u_R^*$                    & $u_L^c$   & $\bar{3}$ & 1 & $-$4/3 \\
        & $\hat D^c$ & $\widetilde d_R^*$                    & $d_L^c$   & $\bar{3}$ & 1 & \hphantom{$-$}2/3 \\ \hline
Higgs boson/   & $\hat H_d$ & $(H^0_d\,,\,H_d^-)$ & $(\widetilde H^0_d,\widetilde H^-_d)$ \vphantom{$\widetilde {H^H}$} & 1 & 2 & $-$1 \\
higgsino & $\hat H_u$ & $(H^+_u\,,\,H^0_u)$ & $(\widetilde H^+_u,\widetilde H^0_u)$ \vphantom{$\widetilde {H^H}$} & 1 & 2 & \hphantom{$-$}1 \\
\end{pdgwidetable}
}

\index{gauge supermultiplet}%
The gauge supermultiplets consist of the gluons and their
gluino fermionic superpartners and the
SU(2)$\times$U(1) gauge bosons and their
gaugino fermionic superpartners.
\index{matter supermultiplet}%
The matter supermultiplets
consist of three generations of left-handed
quarks and leptons and
their scalar superpartners (squarks and sleptons, collectively referred to as sfermions),
\index{sfermion}%
and the corresponding antiparticles.
The Higgs supermultiplets
consist of two complex Higgs doublets, their
higgsino fermionic superpartners, and the
corresponding antiparticles.
The enlarged Higgs sector of the MSSM constitutes the minimal structure
needed to guarantee the cancellation of gauge
anomalies\cite{Georgi:1972bb}
generated by the
higgsino superpartners that can appear as internal lines in triangle diagrams with
three external electroweak gauge bosons.
Moreover, without a second Higgs doublet, one cannot
generate mass for both ``up''-type and ``down''-type
quarks (and charged leptons)
in a way consistent with the underlying
SUSY\cite{Fayet:1974pd,Inoue:1982ej,Gunion:1984yn}.

In the most elegant treatment of SUSY, spacetime is extended to superspace
\index{superspace} which consists
of the spacetime coordinates and new anticommuting fermionic coordinates $\theta$ and $\theta^\dagger$\cite{Salam:1974yz,susy1:WessBagger,susy1:DHM}{}.
Each supermultiplet is represented by a superfield that is a function of the superspace coordinates.
The fields of a given supermultiplet (which are functions of the spacetime coordinates) are
coefficients of the $\theta$ and $\theta^\dagger$ expansion of the corresponding superfield.

Vector superfields 
\index{vector superfields}%
contain the gauge-boson fields and their gaugino partners.  Chiral superfields
\index{chiral superfields}%
contain
the spin-0 and spin-1/2 fields of the matter or Higgs supermultiplets.
A general supersymmetric Lagrangian
is determined by three
functions of the chiral superfields\cite{susy1:Weinberg00}{}:~the
superpotential, the
\index{superpotential}%
K\"ahler potential, and the gauge kinetic
function (which can be appropriately
generalized to accommodate higher derivative terms\cite{Buchbinder:1994iw,*Antoniadis:2007xc,*Dudas:2015vka}{}).
Minimal forms for the K\"ahler potential 
\index{K\"ahler potential}%
and gauge kinetic function, which 
\index{gauge kinetic function}%
generate canonical kinetic energy terms for all the fields,
are required for renormalizable globally supersymmetric theories.
A renormalizable superpotential,
which is at most cubic in the chiral superfields, yields
supersymmetric Yukawa couplings and mass terms.  A combination
of gauge invariance and SUSY produces couplings of gaugino
fields to matter (or Higgs) fields and their corresponding superpartners.
The (renormalizable)
MSSM Lagrangian is then constructed by including all possible
supersymmetric interaction terms (of dimension four or less) that satisfy
SU(3)$\times$SU(2)$\times$U(1) gauge invariance and \hbox{$B\!\!-\!\!L$}
conservation (where $B\!=\!$~baryon number and $L\!=\!$~lepton number).
Finally, the most general soft-super\-symmetry-breaking terms
consistent with these symmetries are added\cite{Girardello:1981wz,Hall:1990ac,*Jack:1999ud,Chung:2003fi}.

Although the MSSM is the focus of much of this review, there is
some motivation for
considering non-minimal supersymmetric extensions of the SM\cite{susy1:Moretti}.
For example, extra
structure is needed to generate non-zero neutrino masses
as discussed in \Sec{susy1:sec:massive-neutral}.
In addition, in order to address some theoretical issues
and tensions associated with the MSSM, it has been fruitful to
introduce one additional singlet Higgs superfield.
The resulting next-to-minimal supersymmetric extension of the Standard
\index{NMSSM, next-to-minimal supersymmetric SM}%
\index{next-to-minimal supersymmetric SM, NMSSM}%
Model (NMSSM)\cite{Ellis:1988er,*Ellwanger:1999ji,*Ellwanger:2009dp,*Maniatis:2009re} is briefly considered in Secs.~\ref{susy1:sec:sp-spectrum}--\ref{susy1:sec:exp-MSSM}
and~\ref{muproblem}.  Finally, one is always free to add additional
fields to the SM along with the corresponding superpartners, as noted in \Sec{susy1:sec:extensions}.  For example, the motivation for adding a color octet chiral superfield~\cite{Fayet:1978qc,Hall:1990hq,*Randall:1992cq} to the MSSM is briefly discussed in \Sec{DiracGauginos}.
However, only certain choices for the
new superfields ({\it e.g.}, the addition of complete SU(5)
multiplets) will preserve the successful gauge coupling unification of the MSSM\@.

\subsection{R-parity and the lightest supersymmetric particle}
\label{susy1:sec:R-par-lsp}

The (renormalizable) SM Lagrangian possesses an accidental
global \hbox{$B\!-\!L$} symmetry due to the fact that $B$ and
$L$-violating operators composed of SM fields must have
dimension $d=5$ or larger\cite{Weinberg:1979sa,*Weinberg:1980bf,*Wilczek:1979hc,*Weldon:1980gi}{}.  Consequently, $B$ and
$L$-violating effects are suppressed by $(M_{\rm EW}/M)^{d-4}$, where
$M$ is the characteristic mass scale of the physics that generates the
corresponding higher-dimensional operators.  Indeed, values of $M\gtrsim 10^{16}$~GeV, corresponding to the grand unification (GUT) scale or larger, may be responsible for the observed
(approximate) stability of the proton and suppression of
neutrino masses.\index{GUT (grand unified theory)}
Unfortunately, these results are not guaranteed in a generic
supersymmetric extension of the SM\@.  For example, it is
possible to construct gauge-invariant supersymmetric dimension-four
$B$ and $L$-violating operators made up of fields of SM
particles and their superpartners.  Such operators, if simultaneously present in the
theory, would typically yield a proton decay rate many orders of magnitude
larger than the current experimental bound.  It is for this reason
that \hbox{$B\!-\!L$} conservation is {\it imposed}\/ on the
supersymmetric Lagrangian when defining the MSSM, which is sufficient
for eliminating all $B$ and $L$-violating operators of dimension
$d\leq 4$.

\index{R-parity@R-parity}%
As a consequence of the \hbox{$B\!-\!L$} symmetry, the MSSM possesses
a multiplicative R-parity invariance, where
${R=(-1)^{3(B\!-\!L)+2S}}$
for a particle of spin $S$\cite{Fayet:1977yc,*Farrar:1978xj}{}.
This implies that all the particles of the SM
have even R-parity, whereas the corresponding
superpartners have odd R-parity.
The conservation of R-parity in scattering
and decay processes has a critical impact on supersymmetric
phenomenology.  For example, any initial state in a scattering
experiment will involve ordinary (R-even) particles.
Consequently, it follows that supersymmetric particles must be
produced in pairs.  In general, these particles are highly unstable
and decay into lighter states.  Moreover, R-parity invariance
also implies that
the LSP is absolutely
stable, and must eventually be produced
at the end of a decay chain initiated by the decay of a heavy unstable
supersymmetric particle.
In order to be consistent with cosmological constraints, a stable LSP
is almost certainly electrically and color neutral\cite{Ellis:1983ew}{}.
Consequently, the LSP in an
\index{R-parity conserving}%
R-parity-conserving theory is weakly
interacting with ordinary matter, \ie, it behaves like a stable heavy
neutrino and will escape collider detectors without being directly
observed.  Thus, the canonical signature for conventional
R-parity-conserving supersymmetric theories is missing (transverse)
momentum, due to the escape of the LSP\@.  Moreover, as noted in \Sec{susy1:sec:Intro}
and reviewed in \Ref{susy1:griest,Steffen:2008qp},
the stability of the 
LSP in R-parity-conserving SUSY
makes it a promising candidate for dark matter.

The possibility of relaxing the R-parity invariance of the MSSM (which
would generate new $B$ and/or $L$-violating interactions) will be
addressed in \Sec{susy1:sec:R-parity-viol-susy}.
However, note that in R-parity violating (RPV) models, the LSP is no longer
stable and thus would not be a viable candidate for the dark
matter (unless its lifetime was significantly longer than the age of the universe). In such scenarios, one must look elsewhere to explain the origin of
dark matter.

\subsection{The goldstino and gravitino}
\label{susy1:sec:golds-grav}

In the MSSM, SUSY breaking is implemented by including the
most general renormalizable soft-SUSY-breaking terms
consistent with the SU(3)$\times$SU(2)$\times$U(1) gauge symmetry and
R-parity invariance.  These terms
parameterize our ignorance of the fundamental mechanism of
super\-symmetry breaking.  If super\-symmetry breaking occurs
spontaneously, then a massless Goldstone fermion
called the goldstino ($\widetilde G_{1/2}$) must exist.  The
goldstino would then be the LSP, and could play an important role
in supersymmetric phenomenology\cite{Fayet:1979qi,*Fayet:1979yb}.

\index{goldstino}%
However, the goldstino degrees of freedom are physical only in models of
spontaneously-broken global SUSY\@.
If SUSY is a local symmetry,
then the theory must incorporate gravity; the resulting theory is called
\index{supergravity}%
supergravity\cite{susy1:Nath,susy1:Supergravitybook,*susy1:sugra,*susy1:sugra2}{}.
In models of spontaneously-broken supergravity, the goldstino is ``absorbed''
by the gravitino ($\widetilde G$),
the spin-3/2 superpartner of the graviton, via the
super-Higgs mechanism\cite{Deser:1977uq,*Cremmer:1978iv}{}.  Consequently,
\index{super-Higgs mechanism}%
the goldstino is removed from the
physical spectrum and the gravitino acquires a mass
(denoted by $m_{3/2}$).  If $m_{3/2}$ is smaller than the mass of the lightest
superpartner of the SM particles, then the gravitino
is the LSP. 
\index{gravitino} %

In processes with center-of-mass energy $E\gg m_{3/2}$,
one can employ
the goldstino--gravitino equivalence theorem\cite{Casalbuoni:1988kv,*Casalbuoni:1988qd,*Maroto:1999vd},
which implies that the interactions of
the helicity $\pm\frac{1}{2}$ gravitino (whose properties
approximate those of the goldstino) dominate those of the
helicity $\pm{\frac{3}{2}}$ gravitino.
The interactions of gravitinos with other light fields can be
described by a low-energy effective Lagrangian that is determined
by fundamental principles\cite{Komargodski:2009rz,*Antoniadis:2010hs,*Ghilencea:2015aph}.

\subsection{Hidden sectors and the structure of SUSY breaking}
\label{susy1:sec:hidden-sect}

\index{hidden sector}%
\index{supersymmetry breaking}%
It is very difficult (perhaps impossible) to construct a realistic model of
spontaneously-broken weak-scale super\-symmetry where the
super\-symmetry breaking arises solely as a consequence of
the interactions of the particles of the
MSSM\@.  A more successful scheme 
posits a theory with at least two
distinct sectors: a visible sector
consisting of the particles of the MSSM\cite{Chung:2003fi} and a
sector where SUSY
breaking is generated. 
It is often (but not always) assumed that particles of the
hidden sector are neutral with respect to the
SM gauge group.  The effects of the hidden sector
super\-symmetry breaking are then transmitted to the
MSSM by some mechanism (often involving the mediation by particles
that comprise an additional messenger
sector).  Two theoretical scenarios that exhibit this structure are
gravity-mediated and gauge-mediated SUSY breaking.

\index{supergravity}%
Supergravity models provide a natural mechanism for
transmitting the SUSY breaking of the hidden sector to the
particle spectrum of the MSSM\@. In models of gravity-mediated
SUSY breaking, gravity is the messenger of
super\-symmetry breaking\cite{susy1:sugramodels,*Barbieri:1982eh,*Ibanez:1982qk,*Nilles:1982dy,*Nilles:1982mp,*Cremmer:1982vy,*Ohta:1982wn,Alvarez-Gaume:1983drc,Hall:1983iz,Soni:1983rm,*Kawamura:1994ys,susy1:bim}{}.
More precisely, super\-symmetry breaking is mediated by effects of
gravitational strength (\ie suppressed by inverse powers of the Planck mass).
The soft-SUSY-breaking parameters with dimensions of mass arise as
model-dependent multiples of the gravitino mass $m_{3/2}$.
In this scenario, $m_{3/2}$ is of
order the electroweak-symmetry-breaking scale, while the gravitino couplings are
roughly gravitational in strength\cite{Nilles:1983ge,Lahanas:1986uc}.\footnote{However,
such a gravitino typically plays no direct role in supersymmetric phenomenology at
colliders (except perhaps indirectly in the case where the gravitino is
the LSP\cite{Feng:2003xh,*Feng:2003uy,*Feng:2004gn}{}).}

Under certain theoretical assumptions
that govern the structure of the K\"ahler potential (the so-called sequestered form
introduced in \Ref{Randall:1998uk}), SUSY breaking is due
entirely to the super-conformal (super-Weyl) anomaly,
\index{super-conformal anomaly}%
which is common to all supergravity models\cite{Randall:1998uk}{}.
In particular, gaugino masses are radiatively generated at one-loop,
and squark and slepton squared-mass matrices are flavor-diagonal.
In sequestered scenarios, sfermion squared-masses arise at
two-loops, which implies that
gluino and sfermion masses are of the same order of magnitude.
This approach is called anomaly-mediated SUSY breaking (AMSB).
\index{AMSB, anomaly-mediated SUSY breaking}%
\index{anomaly-mediated SUSY breaking, AMSB}%
Indeed, anomaly mediation is more generic than originally conceived,
and provides a ubiquitous source of SUSY breaking\cite{DEramo:2012vvz,*DEramo:2013dzi,*deAlwis:2008aq,*deAlwis:2012gr,*Harigaya:2014sfa}.
However in the simplest formulation of AMSB as applied to the MSSM,
the squared-masses of the sleptons are negative (known as the
tachyonic slepton problem).
It may be possible to cure this otherwise fatal flaw in non-minimal extensions of
the MSSM\cite{Jack:2002pn,*Murakami:2003pb,*Kitano:2004zd,*Hodgson:2005en,*Jones:2006re}{}.
Alternatively, one can assert that
anomaly mediation is not the sole source of SUSY breaking in the sfermion
sector.  In non-sequestered scenarios, sfermion squared-masses can arise at
tree-level, in which case squark masses would be parametrically larger than the loop-suppressed gaugino masses\cite{Asai:2007sw}.

In gauge-mediated super\-symmetry breaking,\index{gauge-mediated supersymmetry breaking}
gauge forces transmit
the super\-symmetry breaking to the MSSM\@.
A typical structure of such models involves a hidden sector
where SUSY is broken, a messenger sector consisting of
particles (messengers) with nontrivial SU(3)$\times$SU(2)$\times$U(1) quantum
numbers, and the visible sector consisting of the fields of the
MSSM\cite{Dine:1981za,*Dimopoulos:1981au,*Dimopoulos:1982gm,*Dine:1981gu,*Nappi:1982hm,*AlvarezGaume:1981wy,Dine:1993yw,*Dine:1994vc,Dine:1995ag,Giudice:1998bp}.
The direct coupling of the messengers to the hidden sector generates a
super\-symmetry-breaking spectrum in the messenger sector.
Super\-symmetry
breaking is then transmitted to the MSSM via the virtual exchange of the
messenger fields.  In models of direct gauge mediation, there is no
separate hidden sector.  In particular, the sector in which the
SUSY breaking originates
includes fields that carry nontrivial SM quantum numbers,
which allows for the direct transmission of SUSY breaking to
the MSSM\cite{Poppitz:1996fw,*Murayama:1997pb,*Luty:1997nq,*Agashe:1998wm,*ArkaniHamed:1997jv,*Csaki:2006wi,*Ibe:2007ab}.

In models of gauge-mediated SUSY breaking with a minimal K\"ahler potential,
the gravitino is the LSP\cite{Pagels:1981ke}, as its mass can range from a few eV
(in the case of low SUSY breaking scales) up to a few GeV
(in the case of high SUSY breaking scales).
In particular, the gravitino is a potential dark matter candidate
\index{Dark matter candidates!gravitino}%
(for a review and guide to the literature, see \Ref{Steffen:2008qp}).
The couplings of the helicity $\pm\frac{1}{2}$ components of
$\widetilde G$ to the particles
of the MSSM (which approximate those of the goldstino
as previously noted in \Sec{susy1:sec:golds-grav})
are significantly stronger than gravitational strength and amenable to
experimental collider analyses.

The mass ranges of the gravitino in either gravity-mediated or gauge-mediated
SUSY breaking are further constrained by cosmological
considerations~\cite{PhysRevLett.48.1303,*Kawasaki:2006hm,*Kawasaki:2008qe}. 
In particular, there is a danger of overabundance of
gravitinos as the dark matter or
modifications to the successful predictions of light element abundances if $\widetilde{G}$ decays before nucleosynthesis.
Avoiding these cosmological gravitino problems imposes strong constraints on
gravity-mediated and gauge-mediated
SUSY breaking models. 

The concept of a hidden sector is more general than SUSY\@.
Hidden valley models\cite{Strassler:2006im,*Han:2007ae} posit the existence of
\index{hidden valley models}%
a hidden sector of new particles and interactions that are very
weakly coupled to particles of the SM\@.
The impact of a hidden valley
on supersymmetric phenomenology at colliders can be
significant if the LSP lies in the hidden sector\cite{Strassler:2006qa,*Zurek:2008qg}.

\subsection{SUSY and extra dimensions}
\label{susy1:sec:extra-dims}

\index{Extra Dimensions}%
Approaches to SUSY breaking have also
been developed in the context of theories in which the number of spatial
dimensions is greater than three.  In particular,
a number of SUSY-breaking mechanisms have been proposed that
are inherently extra-dimensional\cite{susy1:tasi02,*susy1:Csaki}{}.  The size of the
extra dimensions can be significantly larger than $M_{\rm P}^{-1}$;
in some cases of order (TeV)$^{-1}$ or even
larger (\eg, see  Ref.~\cite{PDGextra} and \Ref{susy1:Rubakov,*Hewett:2002hv}).

For example, in one approach
the fields of the MSSM live on some brane (a lower-dimensional manifold
\index{Brane}%
embedded in a higher-dimensional spacetime), while the sector of the
theory that breaks SUSY lives on a second spatially-separated
brane.  Two examples of this approach are
AMSB\cite{Randall:1998uk}
and gaugino-mediated SUSY breaking\cite{Chacko:1999hg,*Kaplan:1999ac,*Chacko:1999mi}{}.  In both cases,
SUSY breaking is transmitted through fields that live in the
bulk (the higher-dimensional space between the two branes).
This setup has some features in common with both gravity-mediated
and gauge-mediated SUSY breaking (\eg, hidden and visible
sectors and messengers).

Since a higher dimensional theory must be compactified to four
spacetime dimensions, one can also generate a source of SUSY breaking by employing
boundary conditions on the compactified space that distinguish between
fermions and bosons.  This is the so-called
Scherk-Schwarz mechanism\cite{Scherk:1978ta,*Scherk:1979zr}{}. The
\index{Scherk-Schwarz mechanism}%
phenomenology of such models can be strikingly different from
that of the usual MSSM\cite{Barbieri:2001yz,*Barbieri:2001dm,*Garcia:2015sfa}.

\subsection{Split-SUSY}\label{susy1:sec:split-susy}

\index{split supersymmetry}%
If SUSY is not connected with the origin of the electroweak scale,
it may still be possible that some remnant
of the superparticle spectrum
survives down to the TeV-scale or below.
This is the idea of split-SUSY\cite{Wells:2004di,ArkaniHamed:2004fb,*Giudice:2004tc}{}, in
which scalar superpartners of the quarks and leptons
are significantly heavier than 1~TeV, whereas the fermionic
superpartners of the gauge and Higgs bosons have masses on the order of
1~TeV or below.
With the exception of a single light neutral scalar whose
properties are practically
indistinguishable from those of the SM Higgs
boson, all other Higgs bosons are also assumed to be very heavy.
Among the supersymmetric particles, only the fermionic superpartners
may be kinematically accessible at the LHC.

In models of split SUSY, the top squark masses cannot be arbitrarily large, as these
parameters enter in the radiative corrections to the mass of the observed
Higgs boson\cite{Giudice:2011cg,Arvanitaki:2012ps,*ArkaniHamed:2012gw,Slavich:2020zjv}.
In the MSSM, a Higgs boson mass of
125 GeV (see Ref.~\cite{PDGhiggs}) implies an upper bound on the top squark
mass scale
in the range of 10 to $10^8$~TeV\cite{Bagnaschi:2014rsa,*Vega:2015fna}, depending on the value of
the ratio of the two neutral Higgs field vacuum expectation values (VEVs), although
this mass range can be somewhat extended by varying other
relevant MSSM parameters.
In some approaches, gaugino masses
are one-loop suppressed relative to the sfermion masses, corresponding to the so-called mini-split
SUSY spectrum\cite{Arvanitaki:2012ps,*ArkaniHamed:2012gw,Kahn:2013pfa,*Baer:2024fgd}{}.
The higgsino mass scale may or may not be likewise
suppressed depending on the details of the model\cite{Hall:2011jd,*Ibe:2011aa}{}.

The SUSY breaking
required to produce such a split-SUSY spectrum would destabilize
the gauge hierarchy, and thus would not provide an
explanation for the scale of electroweak symmetry breaking.
Nevertheless, models of split-SUSY
\index{Dark matter candidates!gaugino}%
\index{Dark matter candidates!higgsino}%
can account for the dark matter (which is assumed to be the LSP
gaugino or higgsino) and gauge coupling unification, thereby
preserving two of the desirable features of weak-scale SUSY\@.
Finally, as a consequence of the very large squark and slepton masses,
neutral flavor changing and CP-violating effects, which can be
problematic in models with TeV-scale SUSY-breaking masses,
are sufficiently reduced to avoid conflict with experimental observations.

\section{Parameters of the MSSM}\label{susy1:sec:MSSM-params}

\index{MSSM parameters}%
The parameters of the MSSM are conveniently described by considering
separately the
super\-symmetry-conserving and
the super\-symmetry-breaking sectors.
A careful discussion of the conventions used here in defining the
tree-level MSSM parameters can be found in Refs.~\cite{Haber:2017aci,Allanach:2008qq,susy1:DHM}.
For simplicity, consider first the
case of one generation of quarks, leptons, and their scalar
superpartners.

\subsection{The SUSY-conserving parameters}
\label{susy1:sec:susy-conserv}

The parameters of the super\-symmetry-conserving
sector consist of: (i)~gauge couplings, $g_s$, $g$, and $g^\prime$,
corresponding
to the SM gauge group SU(3)$\times$SU(2)$\times$U(1)
respectively; (ii)~a
super\-symmetry-conserving higgsino mass parameter
$\mu$; and (iii)~Higgs-fermion Yukawa couplings,
$y_u$, $y_d$, and $y_e$,
 of one generation of left- and right-handed
quarks and leptons, and their
superpartners to the Higgs bosons and higgsinos.  Because there is no
right-handed neutrino/sneutrino in the MSSM as defined
here, a Yukawa coupling $y_\nu$ is not included.
The complex $\mu$ parameter and Yukawa couplings
enter via the most general renormalizable R-parity-conserving
superpotential,
\begin{equation}\label{susy1:eq:Wsup}
W_{\rm MSSM}= y_d \hat H_d \hat Q\hat D^c-
y_u \hat H _u\hat Q \hat U^c
+y_e\hat H_d\hat L \hat E^c+\mu \hat H_u \hat H_d\,,
\end{equation}
where the superfields are defined in Table 1 and the
gauge group indices are suppressed. 
\index{mu@$\mu$-term}
More explicitly,
the so-called ``$\mu$-term'' 
can be written out as 
$\mu \epsilon^{mn}(\hat{H}_u)_m (\hat{H}_d)_n$ with an implicit sum over repeated indices, where
$\epsilon^{mn}$ is used to tie together the SU$(2)$ weak isospin
indices $m,n\in\{1,2\}$ in a gauge-invariant way
(where $\epsilon^{12}=-\epsilon^{21}=1$ and $\epsilon^{11}=\epsilon^{22}=0$). 
Likewise, the term $ \hat H _u\hat Q \hat U^c$ can be
written out as $\epsilon^{mn}(\hat{H}_u)_m \hat{Q}_{in}(\hat U^c)^i$,
where there is an implicit sum over the SU(3) color index, $i\in\{1,2,3\}$.
Finally, $\hat H_d \hat Q\hat D^c$ can be written out as  
$\epsilon^{mn}(\hat{H}_d)_m \hat{Q}_{in}(\hat{D^c})^i$, with an
analogous expression for $\hat H_d\hat L \hat E^c$. One can easily generalize \Eq{susy1:eq:Wsup} to a three generation model where $y_u$, $y_d$, and $y_e$ are $3\times 3$ matrices with the corresponding family indices suppressed.

\subsection{The SUSY-breaking parameters}
\label{susy1:sec:susy-break}

The super\-symmetry-breaking
sector contains the following sets of parameters:
(i)~three complex
gaugino Majorana mass parameters, $M_3$, $M_2$, and $M_1$, associated with
the SU(3), SU(2), and U(1) subgroups of the SM;
(ii)~five sfermion squared-mass parameters, $M^2_{\widetilde{Q}}$,
$M^2_{\widetilde{U}}$, $M^2_{\widetilde{D}}$, $M^2_{\widetilde{L}}$, and $M^2_{\widetilde{E}}$,
corresponding to the five electroweak gauge multiplets,
\ie, superpartners of the 
left-handed fields $(u, d)_L$, $u^c_L$,
$d^c_L$, $(\nu$, $e^-)_L$, and $e^c_L$, where the superscript $c$
indicates a charge-conjugated fermion field\cite{susy1:Lang}; and
(iii)~three Higgs-squark-squark and Higgs-slepton-slepton trilinear
interaction terms, with complex coefficients $T_U\equiv y_u A_U$,
$T_D\equiv y_d A_D$, and $T_E\equiv y_e A_E$
\index{A-parameter@$A$-parameter}%
(which define the ``$A$-parameters''), following the notation
employed in \Ref{Allanach:2008qq}.
It is conventional to separate out the
factors of the Yukawa couplings in defining the
$A$-parameters~\cite{Nilles:1983ge,susy1:DHM} (originally motivated by a simple class of
gravity-mediated SUSY-breaking
models~\cite{Nilles:1983ge}).
If the $A$-parameters
are parametrically of the same order (or smaller) relative
to other SUSY-breaking mass parameters, then in most cases
only the third generation $A$-parameters will be
phenomenologically relevant.

Finally, we have
(iv)~two real squared-mass parameters, $m_{H_d}^2$ and~$m_{H_u}^2$ (also
called $m_1^2$ and $m_2^2$, respectively, in the literature), and one
complex squared-mass parameter, 
$m_{12}^2\equiv \mu B$
(the latter defines the ``$B$-parameter''),\footnote{Again, motivated by a simple class of gravity-mediated SUSY-breaking models, it is conventional to separate out the factor of $\mu$ in defining the $B$-parameter~\cite{Nilles:1983ge,susy1:DHM}.  In more general SUSY-breaking scenarios, it is possible to generate $m_{12}^2\neq 0$ even when $\mu=0$.} which appear in the MSSM
tree-level scalar Higgs potential\cite{Gunion:1984yn,susy1:DHM},
\begin{equation}\label{susy1:eq:Hpot}
\begin{aligned}
V&=(m_{H_d}^2+|\mu|^2)H_d^\dagger H_d+(m_{H_u}^2+|\mu|^2)H_u^\dagger
H_u+(m_{12}^2H_u H_d+{\rm h.c.})\\
&   +\frac{1}{8}(g^2+g^{\prime\,2})(H_d^\dagger H_d-H_u^\dagger
H_u)^2+\frac{1}{2}g^2|H_d^\dagger H_u|^2\,,\\
\end{aligned}
\end{equation}
where the SU(2)-invariant combination of the complex doublet scalar
fields $H_u$ and $H_d$ that appears in \Eq{susy1:eq:Hpot} is given by
$H_u H_d\equiv \epsilon^{mn}(H_u)_m(H_d)_n=H_u^+H_d^--H_u^0 H_d^0$.
Note that the quartic Higgs couplings are related to the gauge
couplings $g$ and $g'$ as a consequence of SUSY\@.
The breaking of the
SU(2)$\times$U(1) electroweak symmetry group to U(1)$_{\rm EM}$ is
only possible after incorporating the
SUSY-breaking Higgs squared-mass parameters $m_{H_d}^2$, $m_{H_u}^2$
(which can be negative) and $m_{12}^2$.
After minimizing the Higgs scalar potential,
these three squared-mass
parameters can be re-expressed in terms of the two
Higgs VEVs, $\langle H_d^0\rangle\equiv v_d/\sqrt{2}$
and $\langle H_u^0\rangle\equiv v_u/\sqrt{2}$,
and the CP-odd Higgs mass~$m_A$ [cf.~Eqs. (\ref{susy1:eq:minbeta})
and (\ref{susy1:eq:minconditions}) below].
One is always free to rephase the Higgs doublet fields such that $v_d$
and $v_u$  (also called $v_1$ and
$v_2$, respectively, in the literature) are both real and positive.

The quantity, $v^2\equiv v_d^2+v_u^2=
4m_W^2/g^2=(2G_F^2)^{-1/2}\simeq (246~{\rm GeV})^2$, is fixed by the
\index{Fermi constant}%
Fermi constant, $G_F$, whereas the ratio
\begin{equation}\label{susy1:eq:eqtanbeta}
\tan \beta = v_u/v_d
\end{equation}
is a free parameter such that $0<\beta<\pi/2$.  By employing the
tree-level conditions resulting from the minimization of the scalar
potential, one can eliminate
the diagonal and off-diagonal Higgs squared-masses in favor of $m^2_Z={\frac{1}{4}} (g^2+ g^{\prime\,2})v^2$,
the CP-odd Higgs mass $m_A$ and the parameter $\tan\beta$,
\begin{equation}\label{susy1:eq:minbeta}
\begin{aligned}
\sin 2\beta &= \frac{2m_{12}^2} {m_{H_d}^2+m_{H_u}^2+2|\mu|^2}=\frac{2m_{12}^2}{m_A^2}
\,, \\
\end{aligned}
\end{equation}
\begin{equation}\label{susy1:eq:minconditions}
\begin{aligned}
\frac{1}{2} m_Z^2 &= -|\mu|^2+{\frac{m_{H_d}^2-m_{H_u}^2\tan^2\beta}{\tan^2\beta-1}}\,.
\end{aligned}
\end{equation}
One must also guard against the existence of
charge and/or color breaking global minima
due to non-zero VEVs for the squark and
charged slepton fields.  This possibility can be avoided
if the $A$-parameters are not unduly
large\cite{Alvarez-Gaume:1983drc,Frere:1983ag,*Derendinger:1983bz,*Gunion:1987qv,*Chowdhury:2013dka,*Hollik:2016dcm,Casas:1995pd}.
Additional constraints must also be respected to avoid the
possibility of directions in scalar field space in which
the full tree-level scalar potential can become unbounded from
below\cite{Casas:1995pd}. A computer program has been developed
to calculate vacuum stability bounds in general models at the one-loop
level~\cite{Camargo-Molina:2013qva}, 
and has been applied to the MSSM in \Ref{Blinov:2013fta}.  

Note that SUSY-breaking
mass terms for the fermionic superpartners of the scalar fields and
non-holomorphic trilinear scalar interactions (\ie, interactions that mix
scalar fields and their complex conjugates) have not been included above
in the soft-SUSY-breaking sector.
These terms can potentially destabilize the gauge
hierarchy\cite{Girardello:1981wz} in models
with a gauge-singlet superfield.
The latter is not present in the MSSM; hence
as noted in \Ref{Hall:1990ac,*Jack:1999ud}, these so-called non-standard
soft-SUSY-breaking terms are benign.
The phenomenological impact of non-holomorphic soft SUSY-breaking
terms has been reconsidered in Refs.~\cite{Un:2014afa,Ross:2016pml,Ross:2017kjc}.
However, in the most common approaches to constructing a fundamental
theory of SUSY-breaking, the coefficients of these terms
(which have dimensions of mass)
are significantly suppressed compared to the TeV-scale\cite{Martin:1999hc}{}.
Consequently, we follow the usual approach in the literature and omit these terms from
further consideration.
\subsection{MSSM-124}
\label{susy1:sec:MSSM-124}

\index{MSSM-124}%
The total number of independent physical parameters
that define the MSSM (in its most general form) is
quite large, primarily due to the
soft-super\-symmetry-breaking sector.  In particular, in the case of
three generations of quarks, leptons, and their superpartners,
$M^2_{\widetilde{Q}}$,
$M^2_{\widetilde{U}}$, $M^2_{\widetilde{D}}$, $M^2_{\widetilde{L}}$, and $M^2_{\widetilde{E}}$
are hermitian $3\times 3$ matrices, and
$A_U$, $A_D$, and $A_E$ are complex $3\times 3$
matrices.  In addition, $M_1$, $M_2$, $M_3$, $B$, and $\mu$
are in general complex parameters.  Finally, as in the SM, the
Higgs-fermion Yukawa couplings, $y_f$ ($f\!=\!u$, $d$, and $e$),
are complex $3\times 3$ matrices that
are related to the quark and lepton mass matrices via: $M_f= y_f
v_f/\sqrt{2}$, where $v_e= v_d$ [with $v_u$ and $v_d$ as defined
above \Eq{susy1:eq:eqtanbeta}].

However, not all these parameters are physical.
Some of the MSSM parameters can be eliminated by
expressing interaction eigenstates in terms of the mass eigenstates,
with an appropriate redefinition of the MSSM fields to remove unphysical
degrees of freedom.  The analysis of \Ref{Dimopoulos:1995ju,*Sutter:1995kp} shows that the MSSM
possesses 124 independent real degrees of freedom.  Of these, 18
correspond to SM parameters
(including the QCD vacuum angle $\theta_{\rm QCD}$), one corresponds to
a Higgs sector parameter (the analogue of the SM
Higgs mass), and 105 are genuinely new parameters of the model.
The latter include: five real parameters and three CP-violating phases in
the gaugino/higgsino sector, 21 squark and slepton (sfermion) masses,
36 real mixing angles to define the
sfermion mass eigenstates, and 40 CP-violating phases that
can appear in sfermion interactions.
The most general parameterization of the R-parity-conserving MSSM (without additional theoretical
assumptions) will be denoted henceforth as MSSM-124\cite{Haber:1997if}.

\section{The supersymmetric-particle spectrum}\label{susy1:sec:sp-spectrum}

\index{supersymmetric particle spectrum}%
The supersymmetric particles (sparticles) 
\index{sparticle}%
differ in spin by half a unit from their SM partners.
The superpartners of the gauge and Higgs bosons are fermions,
whose names are obtained by appending ``ino'' to the end of the
corresponding SM particle name.  The gluino is the
color-octet Majorana fermion partner of the gluon
with mass $M_{\widetilde g}=|M_3|$.
The superpartners of the electroweak gauge
and Higgs bosons (the gauginos and higgsinos)
can mix due to SU(2)$\times$U(1) breaking effects.  As a result,
the physical states of definite mass are parameter-dependent linear combinations
of the charged or neutral gauginos and higgsinos,
called charginos and neutralinos, respectively
(sometimes collectively called electroweakinos).
The neutralinos are Majorana fermions, which can lead to some
distinctive phenomenological signatures\cite{Barnett:1993ea,*Baer:1989hr,Bilenky:1985wu,*Bilenky:1986nd,*MoortgatPick:2002iq}{}.
The superpartners of the quarks and leptons are spin-zero
bosons, with an ``s'' appended to the beginning of the corresponding SM particle name: the squarks, charged sleptons,
and sneutrinos, respectively.
A complete set of Feynman rules for the sparticles of the MSSM can be found in
\Ref{Rosiek:1989rs,*Kuroda:1999ks}.  The MSSM Feynman rules are also implicitly contained in
several amplitude generation and Feynman diagram software packages
(\eg, see Refs. \cite{Alwall:2007st,Hahn:2000kx,Pukhov:1999gg,*CompHEP:2004qpa,*susy1:comphep-sr1}).

It should be noted that all mass formulae quoted below in this Section are
tree-level results.
Radiative loop corrections will modify these results and
must be included in any precision study of supersymmetric
phenomenology\cite{Pierce:1996zz}{}.  Beyond tree level, the definition of
the supersymmetric parameters becomes convention-dependent.  For
example, one can define physical couplings or running couplings, which
differ beyond the tree level.
This provides a challenge to any effort
that attempts to extract supersymmetric parameters from data.
The SUSY Les Houches Accord
(SLHA)\cite{Allanach:2008qq,Skands:2003cj}
\index{SLHA, SUSY Les Houches accord}%
\index{SUSY Les Houches accord, SLHA}%
has been adopted, which establishes
a set of conventions for specifying generic file structures for
supersymmetric model specifications and input parameters,
supersymmetric mass and coupling spectra, and decay tables.  These
provide a universal interface between spectrum calculation programs,
decay packages, and high energy physics event generators.

\subsection{The charginos and neutralinos}
\label{susy1:sec:charginos-neut}

\index{charginos}%
\index{neutralinos}%
\index{gaugino}%

The mixing of the charged gauginos ($\widetilde W^\pm$) and charged
higgsinos ($\widetilde H_u^+$ and $\widetilde H_d^-$) is described (at tree-level)
by a $2\times 2$ complex
mass matrix\cite{susy1:Explicitforms,Kneur:1998gy},
\begin{equation}\label{susy1:eq:charginomatrix}
\begin{aligned}
M_C\equiv 
 \begin{pmatrix}
    M_2\quad
      & \frac{1}{\sqrt{2}} gv_u \\
       \frac{1}{\sqrt{2}} gv_d    \quad
      &\mu \\
\end{pmatrix}
  \,.
\end{aligned}
\end{equation}
To determine the physical chargino states and their
masses,
one must perform a singular value decomposition\cite{susy1:horn}
of the complex matrix $M_C$~\cite{susy1:DHM,Dreiner:2008tw}:
\begin{equation}\label{susy1:eq:svd}
U^* M_C V^{-1}={\rm diag}(M_{\chinoonep}\,,\,M_{\chinotwop})\,,
\end{equation}
where $U$ and $V$ are unitary matrices, and the right-hand side of
\Eq{susy1:eq:svd} is the diagonal matrix of (real non-negative) chargino masses.
Explicit formulae for the singular value decomposition of $M_C$ can be found
in \Ref{Haber:2020wco}.
The physical chargino states are denoted by
$\chinoonepm$ and $\chinotwopm$.  These are linear combinations of the
charged gaugino and higgsino states determined
by the matrix elements of $U$ and $V$\cite{susy1:Explicitforms,Kneur:1998gy}
The chargino masses correspond to the singular values\cite{susy1:horn} of
$M_C$, \ie, the positive square roots
of the eigenvalues of $M_C^\dagger M_C$:
\begin{equation} \label{susy1:eq:chimasses}
\begin{aligned}
&M^2_{\chinoonep,\chinotwop}=
\frac{1}{2}\biggl\{|\mu|^2+|M_2|^2+2m_W^2
\bookcr{}{&}\mp\sqrt{\left(|\mu|^2+|M_2|^2+2m_W^2\right)^2 
-4 |\mu M_2 - m_W^2 \sin2\beta|^2}\,\,
\biggr\}\,,
\end{aligned}
\end{equation}
in a convention where $v_u$ and $v_d$ are real and positive, and where the states are
ordered such that $M_{\chinoonep} \leq M_{\chinotwop}$. 
The relative phase of $\mu^*$ and $M_2$ is physical and potentially
observable~\cite{Pokorski:1999hz}. 

The mixing of the neutral gauginos ($\widetilde B$ and
$\widetilde W^0$) and neutral
higgsinos ($\widetilde H_d^0$ and $\widetilde H_u^0$) is
described (at tree-level) by a $4\times 4$ complex symmetric mass
matrix\cite{susy1:Explicitforms,Kneur:1998gy},
\begin{equation}\label{susy1:eq:neutralinomatrix}
M_N\equiv 
\begin{pmatrix}
    M_1\quad & 0 \quad & -\frac{1}{2} g' v_d \quad & \phantom{-}\frac{1}{2} g'
    v_u \\[4pt]
 0 \quad & M_2 \quad & \!\!\phantom{-}\frac{1}{2} g v_d \quad & -\frac{1}{2} g v_u \\[4pt]
-\frac{1}{2} g' v_d \quad & \phantom{-}\frac{1}{2} g v_d \quad & 0 \quad & -\mu \\[4pt]
\phantom{-}\frac{1}{2} g' v_u \quad & -\frac{1}{2} g v_u \quad & -\mu \quad & 0 
\end{pmatrix}  \,.
\end{equation}
To determine the physical neutralino states and their masses,
one must perform an
Autonne-Takagi factorization \cite{susy1:autonne,*susy1:takagi,susy1:horn}
(also called Takagi diagonalization in Refs.~\cite{susy1:DHM,Choi:2006fz,Dreiner:2008tw})
of the complex symmetric matrix $M_N$:
\begin{equation}\label{susy1:eq:takagi}
W^T M_N W={\rm diag}(M_{\ninoone}\,,\,M_{\ninotwo}\,,\,M_{\ninothree}\,,\,M_{\ninofour})\,,
\end{equation}
where $W$ is a unitary matrix (which is called $N^{-1}$ in
Refs.~\cite{Haber:1984rc,Gunion:1984yn}) and the right-hand side of
\Eq{susy1:eq:takagi} is the diagonal matrix of (real non-negative) neutralino masses.
The physical neutralino states are denoted by
$\widetilde\chi_i^0$ (for $i=1,\ldots, 4$), where the states are ordered such that
$M_{\ninoone}\leq M_{\ninotwo}\leq M_{\ninothree}\leq M_{\ninofour}$.
The $\widetilde\chi_i^0$ are the linear combinations of the
neutral gaugino and higgsino states determined
by the matrix elements of $W$.
The neutralino masses correspond to the singular values of
$M_N$, \ie, the positive square roots
of the eigenvalues of $M_N^\dagger M_N$.  Exact formulae for these
masses can be found in \Ref{Choi:2001ww,*ElKheishen:1992yv}.  A numerical
algorithm for determining the mixing matrix $W$ has been given in
\Ref{Hahn:2006hr}.

If a chargino or neutralino state approximates a particular gaugino or
higgsino state, it is convenient to employ the corresponding
nomenclature.  Specifically, if $|M_1|$ and $|M_2|$ are small compared to
$m_Z$ and $|\mu|$, then the lightest neutralino $\widetilde\chi_1^0$
would be nearly a pure photino,
$\widetilde\gamma$, the superpartner of the photon.
If $|M_1|$ and $\mZ$ are small compared to $|M_2|$ and
$|\mu|$, then the lightest neutralino would be nearly a pure
bino, $\widetilde B$, the superpartner of the
weak hypercharge gauge boson.  If $|M_2|$ and $\mZ$ are small compared to
\index{bino}%
$|M_1|$ and $|\mu|$, then the lightest chargino pair
and neutralino would constitute
a triplet of roughly mass-degenerate pure
winos, $\widetilde W^\pm$, and $\widetilde W_3^0$,
\index{wino}%
the superpartners of the
weak SU(2) gauge bosons.  Finally, if $|\mu|$ and $\mZ$ are
small compared to $|M_1|$
and $|M_2|$, then the lightest chargino pair and neutralino
would be nearly pure
higgsino states, the superpartners of the Higgs bosons.
Each of the above cases leads to a strikingly different
phenomenology.

In the NMSSM, an additional Higgs singlet superfield is added
to the MSSM\@.  This superfield comprises two real Higgs scalar degrees of freedom
and an associated neutral higgsino degree of freedom.  Consequently,
there are five neutralino mass eigenstates that are obtained by a
Takagi-diagonalization of the $5\times 5$ neutralino mass matrix.  In many
cases, the fifth neutralino state is dominated by its SU(2)$\times$U(1) singlet
component, and thus is very weakly coupled to the SM particles and
their superpartners.

\subsection{The squarks and sleptons}\label{susy1:sec:squarks-slep}

\index{squarks}%
\index{sleptons}%
\index{sneutrinos}%

For a given Dirac fermion $f$, there are two superpartners, $\widetilde
f_L$ and $\widetilde f_R$, where the $L$ and R subscripts simply identify
the scalar partners that are related by SUSY to the left-handed and
right-handed fermions, $f_{L,R}$, as indicated in Table~\ref{susy1:tab:MSSM}.
(Since right-handed neutrinos lie outside the SM, there is no corresponding $\widetilde\nu_R$ in the MSSM\@.)
However, $\widetilde f_L$--$\widetilde f_R$ mixing is possible,
in which case $\widetilde f_L$ and $\widetilde f_R$ are not mass
eigenstates.  For three generations of squarks, one
must diagonalize $6\times 6$ matrices corresponding
to the basis $(\widetilde q_{iL}, \widetilde q_{iR})$,
where $i=1,2,3$ are the generation
labels.
For simplicity, only the one-generation case is illustrated
in detail below.

Using the notation of the third family, the one-generation
tree-level squark squared-mass matrix is given by\cite{susy1:DHM,Ellis:1983ed,*Browning:2000hk,*Bartl:2003he,*Bartl:2003pd},
\begin{equation}\label{susy1:eq:squarkmatrix}
{\cal M}^2 = 
\begin{pmatrix}
    M^2_{\widetilde Q}+ m^2_q+ L_q\quad
      & m_q X_q^* \\
    m_q X_q\quad
      &M^2_{\widetilde R}+ m^2_q+ R_q \\
\end{pmatrix}
  \,,
\end{equation}
where
\begin{equation}\label{susy1:eq:xfdef}
X_q\equiv A_q-\mu^* (\cot\beta)^{2T_{3q}}\,,
\end{equation}
and $T_{3q}=\frac{1}{2}$ [$-\frac{1}{2}$] for $q=t$ [$b$].
The diagonal squared-masses are governed by soft-SUSY-breaking
squared-masses $M^2_{\widetilde Q}$ and $M^2_{\widetilde R}\equiv
M^2_{\widetilde U}$ [$M^2_{\widetilde D}$] for $q=t$~[$b$], the
corresponding quark masses $m_t$ [$m_b$] and the electroweak correction terms:
\begin{equation}\label{susy1:eq:sfermion}
\begin{aligned}
L_q &\equiv
(T_{3q}-e_q\sin^2\theta_W)m_Z^2\cos 2\beta\,,\\
R_q &\equiv
e_q\sin^2\theta_W \,m_Z^2\cos 2\beta\,,
\end{aligned}
\end{equation}
where $e_q=\frac{2}{3}$ [$-\frac{1}{3}$] for $q=t$ [$b$].
The off-diagonal squark squared-masses are
proportional to the corresponding quark masses and depend on
$\tan\beta$, the
soft-SUSY-breaking $A$-parameters and the higgsino mass parameter
$\mu$.
Assuming that the $A$-parameters
are parametrically of the same order (or smaller) relative
to other SUSY-breaking mass parameters, it then follows that the first and
second generation 
$\widetilde q_L$--$\widetilde q_R$ mixing
is
smaller than that of the third generation
where mixing can be enhanced by factors of $m_t$ and $m_b\tan\beta$.

In the case of third generation $\widetilde q_L$--$\widetilde q_R$
mixing ($q=t$ or $b$), the top squark 
\index{stop (top squark)} (also called \textit{stop}) and bottom squark (also called \textit{sbottom}) mass 
\index{sbottom (bottom squark)} eigenstates, denoted by $\widetilde q_1$ and
$\widetilde q_2$ (with $m_{\tilde q_1}<m_{\tilde q_2}$), are determined
by diagonalizing the $2\times 2$ matrix ${\cal M}^2$ given by
\Eq{susy1:eq:squarkmatrix}.  The corresponding squared-masses
and mixing angle are given by\cite{Ellis:1983ed,*Browning:2000hk,*Bartl:2003he}:
\begin{equation}\label{susy1:eq:sfmix}
\begin{aligned}
  m^2_{\tilde q_{1,2}} &={\frac{1}{2}}\left[{\rm Tr}\,{\cal M}^2\mp
\sqrt{({\rm Tr}{\cal M}^2)^{2}
-4\,{\rm det}\,{\cal M}^2}\right]\,, \\
\noalign{\smallskip}%
\sin 2\theta_{\tilde q} &= {\frac{2m_q |X_q|}{m^2_{\tilde
q_2}-m^2_{\tilde q_1}}}\,.
\end{aligned}
\end{equation}
The above results 
also apply to the charged sleptons, with the obvious
substitutions: $q\to \ell$ with
$T_{3\ell}=-\frac{1}{2}$ and $e_\ell=-1$, and the
replacement of the SUSY-breaking parameters:
$M^2_{\widetilde Q}\to M^2_{\widetilde L}$,
$M^2_{\widetilde D}\to M^2_{\widetilde E}$, and $A_q\to A_\ell$.
For the neutral sleptons, $\widetilde\nu_R$ does not exist in the
MSSM, so $\widetilde\nu_L$ is a mass eigenstate.

In the case of three generations, the SUSY-breaking scalar-squared
masses [$M_{\widetilde{Q}}^2$, $M_{\widetilde{U}}^2$, $M_{\widetilde{D}}^2$,
$M_{\widetilde{L}}^2$, and $M_{\widetilde{E}}^2$] and
the $A$-parameters [$A_U$, $A_D$, and $A_E$]
are now $3\times 3$ matrices as noted in \Sec{susy1:sec:MSSM-124}.
The diagonalization of the $6\times 6$ squark mass
matrices yields $\widetilde f_{iL}$--$\widetilde f_{jR}$
mixing.
In practice, since the
$\widetilde f_L$--$\widetilde f_R$ mixing is appreciable only for the
third generation, this additional complication can often
be neglected (although see \Ref{Hikasa:1987db,*Gabbiani:1988rb,*Brax:1995up} for examples in which the
mixing between the second and third
generation squarks is relevant).

\section{The supersymmetric Higgs sector}\label{susy1:sec:susy-higgs}

Consider first the Higgs
sector of the MSSM\cite{Inoue:1982ej,Gunion:1984yn,susy1:Gunion90,*Djouadi:2005gj,Carena:2002es}{}.
Despite the large number of possible CP-violating phases among the
MSSM-124 parameters, the tree-level MSSM Higgs potential given by \Eq{susy1:eq:Hpot} is
automatically CP-conserving.  This follows from the fact that
the only potentially complex parameter ($m_{12}^2$) of the MSSM Higgs potential
can be chosen real and positive by 
rephasing the Higgs fields, in which case
$\tan\beta$ is a real positive parameter.
Consequently, in the tree-level approximation the physical neutral Higgs scalars are CP-eigenstates.
The MSSM Higgs sector contains five physical spin-zero
particles: a charged Higgs
boson pair ($\Hpm$), two CP-even neutral Higgs bosons (denoted by $h$
and $H$ where $m_{h} < m_{H}$), and one CP-odd neutral
Higgs boson (denoted by $A$).  
In principle, either $h$ or $H$ could be identified with the Higgs boson that was discovered at the LHC~\cite{Bernon:2015qea,Bernon:2015wef}.  Studies of the MSSM parameter space suggest~\cite{Bechtle:2016kui,*Bahl:2018zmf} that it is unlikely that $H$ is the LHC-observed Higgs boson (although this possibility is not yet completely ruled out).  Henceforth, we shall identify $h$ with the observed Higgs boson with $m_h\simeq 125$~GeV\cite{ATLAS:2024fkg,*CMS:2024eka}.

In the NMSSM\cite{Ellis:1988er,*Ellwanger:1999ji,*Ellwanger:2009dp,*Maniatis:2009re}, the scalar component of the singlet Higgs superfield
adds two additional neutral states to the Higgs sector.  In this
model, the tree-level Higgs sector can exhibit explicit CP-violation.
If CP is conserved, then the two extra neutral scalar states are
CP-even and CP-odd, respectively.  These states can potentially mix
with the neutral Higgs states of the MSSM\@.  If scalar states exist
that are dominantly singlet, then they are weakly coupled to
SM gauge bosons and fermions through their small mixing
with the MSSM Higgs scalars.  Consequently, it is possible that one (or
both) of the singlet-dominated states is considerably lighter than the
Higgs boson that was observed at the LHC.

\subsection{The tree-level Higgs sector}\label{susy1:sec:tree-level-higgs}

The tree-level properties of the Higgs sector are
determined by the Higgs potential given by \Eq{susy1:eq:Hpot} and the Yukawa Lagrangian discussed below.  The quartic interaction terms are
manifestly supersymmetric (although these are modified by
SUSY-breaking effects at the loop level).
In general, the quartic couplings arise from two sources:
(i) the supersymmetric generalization of the scalar potential
(the so-called ``$F$-terms''), and (ii) interaction
terms related by SUSY
to the coupling of the scalar fields and the gauge fields, whose
coefficients are proportional to the corresponding
gauge couplings (the so-called ``$D$-terms'').

In the MSSM,
$F$-term contributions to the quartic Higgs self-couplings
are absent.  As a result, the strengths of the MSSM quartic Higgs interactions
are fixed in terms of the gauge couplings, as noted below \Eq{susy1:eq:Hpot}.  Consequently,
all the tree-level MSSM Higgs-sector parameters depend only on
two quantities:
$\tan\beta$ [defined in \Eq{susy1:eq:eqtanbeta}] and one Higgs mass usually
taken to be $m_A$.   For example, the tree-level squared mass of the charged Higgs boson is given by
\begin{equation}
m^2_{H^\pm}=m_A^2+m_W^2\,,
\end{equation}
where $m_A^2=m^2_{H_d}+m^2_{H_u}+2|\mu|^2$ [cf.\ \Eq{susy1:eq:minbeta}] and
\begin{equation}
H^\pm=H_d^\pm\sin\beta+ H_u^\pm\cos\beta\,,\qquad 
A= \sqrt{2}\left({\rm Im\,}H_d^0\sin\beta+{\rm Im\,}H_u^0\cos\beta\right)\,.
\end{equation}
The CP-even scalar mass eigenstate fields $h$ and $H$ are identified by diagonalizing
the $2\times 2$ squared-mass matrix
\begin{equation}  \label{CPevenSqMass}
\mathcal{M}^2 = 
\begin{pmatrix}
\phantom{-}m_A^2 \sin^2\beta + m^2_Z \cos^2\beta \ \ & \quad
           -(m_A^2+m^2_Z)\sin\beta\cos\beta \\[4pt]
  -(m_A^2+m^2_Z)\sin\beta\cos\beta \ \   & \quad
\phantom{-} m_A^2\cos^2\beta+ m^2_Z \sin^2\beta \end{pmatrix}\,.
\end{equation}
In particular,
\begin{align}
h &= -(\sqrt{2}\,{\rm Re\,}H_d^0-v_d)\sin\alpha+
(\sqrt{2}\,{\rm Re\,}H_u^0-v_u)\cos\alpha\,,\\
H &= (\sqrt{2}\,{\rm Re\,}H_d^0-v_d)\cos\alpha+
(\sqrt{2}\,{\rm Re\,}H_u^0-v_u)\sin\alpha\,,
\end{align}
with corresponding tree-level squared masses,
\begin{equation} \label{susy1:eq:treelevelHmass}
  m^2_{H,h} = \frac12 \left( m_A^2 + m^2_Z \pm
                  \sqrt{(m_A^2+m^2_Z)^2 - 4m^2_Z m_A^2 \cos^2 2\beta}
                  \; \right)\,,
\end{equation}
and mixing angle $\alpha$ given by
\begin{equation} \label{susy1:eq:treelevelangle}
\cos\alpha=\sqrt{\frac{m_A^2\sin^2\beta+m_Z^2\cos^2\beta-m_h^2}{m_H^2-m_h^2}}\,,
\end{equation}
in a convention where $|\alpha|\leq\pi/2$, where $m_h^2$ and $m_H^2$ are given by \Eq{susy1:eq:treelevelHmass}.  However, because the off-diagonal elements of $\mathcal{M}^2$ are negative, 
it follows that $-\pi/2\leq\alpha\leq 0$~\cite{Haber:2020wco}.  
In light of \Eq{susy1:eq:treelevelHmass}, the tree-level mass of the 
lighter CP-even Higgs boson is bounded~\cite{Inoue:1982ej,Gunion:1984yn},
\begin{equation}
m_{h}\leq m_Z|\cos 2\beta|\leq m_Z\,.
\end{equation}
This bound can be significantly modified when radiative corrections are
included (see \Sec{susy1:sec:rad-correct-higgs}).

The tree-level Higgs couplings to gauge bosons and the Higgs boson self-couplings are governed by the electroweak gauge couplings and the parameter $\cos(\beta-\alpha)$.  Explicitly,
\begin{equation} \label{cbma}
\cos(\beta-\alpha)=\frac{m_Z^2\sin 2\beta\cos
  2\beta}{\sqrt{(m_H^2-m_h^2)(m_H^2-m_Z^2\cos^2
    2\beta)}}\,.
\end{equation}
Note that $\cos(\beta-\alpha)\to 0$ in the limit of $m_H\gg m_h$, $m_Z$.  In this \textit{decoupling limit}~\cite{Gunion:2002zf},
the properties of $h$ coincide with those of the SM Higgs boson (see \Sec{alignment}).

The tree-level Higgs-quark and Higgs-lepton interactions of the MSSM
are derived from the superpotential given in \Eq{susy1:eq:Wsup}.  The corresponding
Higgs-fermion Yukawa couplings can be expressed in terms of the fermion masses and
the separate parameters $\cos(\beta-\alpha)$ and $\tan\beta$.
In particular, the Higgs sector of
the MSSM is a Type-II two-Higgs doublet model\cite{Hall:1981bc}{}, in which
one Higgs doublet ($H_d$) couples exclusively to the right-handed
down-type quark (or lepton) fields and the second Higgs doublet
($H_u$) couples exclusively to the right-handed up-type quark fields.
For example, the Yukawa Lagrangian that governs the couplings of the Higgs bosons to the third generation of quarks is
\begin{equation} \label{LYUK}
    -\mathcal{L}_Y=\epsilon^{mn}\bigl[y_b \bar{b}_R(H_d)_m (Q_{L3})_n-y_t \bar{t}_R(H_u)_m(Q_{L3})_n\bigr]+{\rm h.c.},
\end{equation}
with an implicit sum over the SU(2)$_L$ indices $m$ and $n$, 
where $Q_{L3}\equiv (t_L\,,\,b_L)$,
$y_b\equiv (y_d)_{33}$ and $y_t\equiv (y_u)_{33}$.  Moreover, after generalizing \Eq{LYUK} to three generations of quarks and charged leptons, 
the diagonalization of the up-type and down-type fermion mass matrices
simultaneously diagonalizes the corresponding Yukawa coupling matrices, resulting in
flavor-diagonal tree-level couplings of the neutral Higgs bosons $h^0$, $H^0$ and
$A^0$ to quark and lepton pairs.  As expected, in the decoupling limit
where $\cos(\beta-\alpha)\to 0$, the couplings of $h$ reduce to those of the SM.

In the NMSSM, we set $\mu=0$ in \Eq{susy1:eq:Wsup} and add two additional terms to the superpotential,
\begin{equation}\label{susy1:eq:WsupN}
W_{\rm NMSSM}\supset \lambda\hat{H}_u\hat{H}_d\hat{S}+\frac{1}{3}\kappa\hat{S}^3\,,
\end{equation}
where $\hat{S}$ is a singlet Higgs superfield.
In the NMSSM as defined here, all terms in $W_{\rm NMSSM}$ are cubic in the
superfields due to the presence of a discrete $\mathbb{Z}_3$ symmetry.
An effective $\mu$-term is generated, $\mu_{\rm eff}=\lambda\langle{S}\rangle$, where $\langle{S}\rangle$ is the VEV of the scalar field component of $\hat{S}$.  Moreover, due to the term proportional to $\lambda$ in \Eq{susy1:eq:WsupN}, there
is now an $F$-term contribution to the quartic Higgs self-couplings. \index{mu@$\mu$-term}
Consequently,
the tree-level bound for the mass of the lightest CP-even MSSM Higgs boson
is modified\cite{Haber:1986gz},
\begin{equation}\label{susy1:eq:nmssmmass}
m_h^2\leq m_Z^2\cos^2 2\beta+\frac{1}{2}\lambda^2 v^2\sin^2 2\beta\,,
\end{equation}
where $v\equiv (v_u^2+v_d^2)^{1/2}=246$~GeV.  By requiring 
that $\lambda$ remain finite after renormalization-group evolution up to
the Planck scale, one finds that $\lambda$ is constrained to lie below about 0.7--0.8 at the electroweak
scale\cite{Ellis:1988er,*Ellwanger:1999ji,*Ellwanger:2009dp,*Maniatis:2009re}{}
(although larger
values of $\lambda$ have also been considered in \Ref{Hall:2011aa}).

\subsection{The radiatively-corrected Higgs sector}\label{susy1:sec:rad-correct-higgs}

When radiative corrections
are incorporated, additional parameters of the supersymmetric model
enter via virtual supersymmetric particles that appear in
loops.  The impact of these corrections
can be significant\cite{Haber:1990aw,*Okada:1990vk,*Ellis:1990nz}{}.
The qualitative behavior of these radiative corrections can be most easily
seen in the large top-squark mass limit.
In addition, we shall assume that both the
splitting of the two diagonal entries and the off-diagonal entries
of the top-squark squared-mass matrix [\Eq{susy1:eq:squarkmatrix}]
are small in comparison to
the geometric mean of the two top-squark squared-masses,
$\msusyy\equiv\mstopa\mstopb$.
In this case (assuming $m_{A}>\mZ$), the predicted upper bound for $m_h$
is approximately given by~\cite{Haber:1996fp}
\begin{equation}\label{susy1:eq:eqmH1up0}
m_{h}^2\lsim m_{Z}^2\cos^2 2\beta+{\frac{3g^2 m_{t}^4}{8\pi^2 m_{W}^2}}\left[\ln\left({\frac{\msusyy}{m_{t}^2}}\right)+
{\frac{X_t^2}{\msusyy}}
\left(1-{\frac{X_t^2}{12\msusyy}}\right)\right],
\end{equation}
where $X_t\equiv A_t-\mu\cot\beta$ [cf.~\Eq{susy1:eq:xfdef}] is proportional to the off-diagonal
entry of the top-squark squared-mass matrix
(where for simplicity, $A_t$ and $\mu$ are taken to be real).
The Higgs mass upper limit specified by \Eq{susy1:eq:eqmH1up0} is saturated when
$\tan\beta$ is large ({\it i.e.}, $\cos^2 2\beta\sim 1$) and $X_t=\sqrt{6}\,
M_S$, which defines the so-called maximal mixing scenario. 
\index{maximal mixing scenario}%

\begin{pdgxfigure}[wide=true, webwidth=0.45\linewidth,  bookwidth = 0.9\linewidth]
\includegraphics{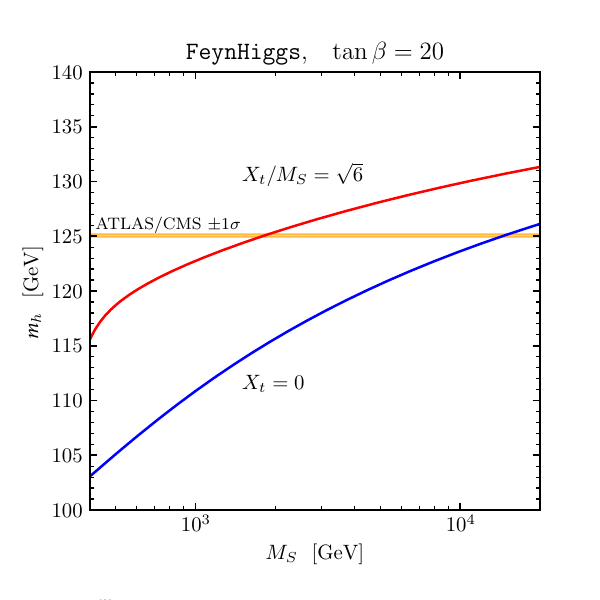}
\includegraphics{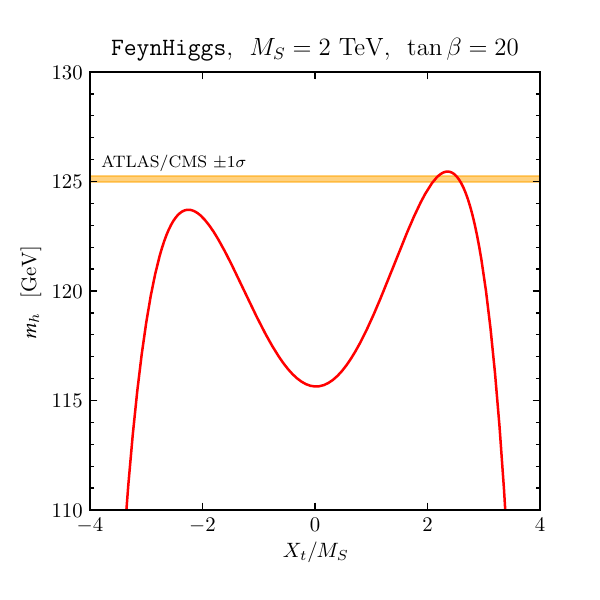}
\caption{The radiatively-corrected value of the MSSM Higgs mass, $m_h$,
as a function of a common
 SUSY mass parameter $M_S$ and the stop mixing parameter $X_t$ (normalized to $M_S$), for $\tan\beta=20$.  The value of the observed Higgs mass currently measured by the ATLAS and CMS collaborations~\protect\cite{ATLAS:2024fkg,*CMS:2024eka} at the LHC is also shown.
This figure has been adapted from Ref.~\protect\cite{Slavich:2020zjv}.}
\label{susy1:fig:kuts}
\end{pdgxfigure}

The set of approximations used to obtain \Eq{susy1:eq:eqmH1up0} somewhat overestimates the value of $\mh$.
A more complete treatment of the MSSM Higgs mass radiative corrections,
which incorporate
renormalization group improvement, two-loop and
leading three-loop contributions~\cite{Draper:2016pys,Bagnaschi:2017xid,Slavich:2020zjv},
yields a predicted value of $\mh$ shown in
Fig.~\ref{susy1:fig:kuts}, as a function of $X_t$ (assumed for simplicity to
be real).  Higher-order radiative corrections beyond those reviewed in Ref.~\cite{Slavich:2020zjv}
have recently been obtained in Ref.~\cite{Bahl:2023ead,*Kwasnitza:2025mge}, although their impact on the results shown in Fig.~\ref{susy1:fig:kuts} is quite small.

The predicted value of \mh can now be used to derive an upper bound for the lightest stop mass in light of the experimentally measured value of \mh{}. Using the \mh measurement along with the requirement of a valid electroweak minimum of the scalar potential, one obtains a rough upper bound of $m_{{\tilde t}_1} \lsim 10^{11}$ GeV~\cite{Allanach:2018fif}. In contrast, since $\mh$ receives radiative corrections from many MSSM fields, obtaining a nontrivial lower bound based on any single parameter such as $m_{{\tilde t}_1}$ would typically require additional assumptions regarding the other MSSM parameters.

The radiative corrections to the scalar squared masses discussed above derive from loop corrections to the tree-level neutral CP-even Higgs squared-mass matrix $\mathcal{M}^2$.  A complete treatment of Higgs sector radiative corrections must also include radiatively generated vertex corrections to the Higgs-fermion Yukawa couplings.  In addition to loop corrections to the Yukawa Lagrangian given in \Eq{LYUK}, new terms in the Yukawa Lagrangian [the so-called wrong-Higgs Yukawa terms~\cite{Haber:2007dj}, \index{wrong-Higgs couplings}
\begin{equation}
-\Delta\mathcal{L}_Y=(\Delta y_b)\bar{b}_R(H_u^*)_m(Q_{L3})_m+(\Delta y_t)\bar{t}_R(H_d^*)_m(Q_{L3})_m+{\rm h.c.},
\end{equation}
are generated due to supersymmetry breaking effects.  The most significant consequence of the wrong-Higgs Yukawa terms is to shift the relation between the bottom quark mass and its corresponding Yukawa coupling~\cite{Hall:1993gn,*Hempfling:1993kv,*Carena:1994bv}
\begin{equation} \label{Deltab}
m_b=\frac{y_b v}{\sqrt{2}}\cos\beta(1+\Delta_b)\,,
\end{equation}
where $\Delta_b\simeq (\Delta y_b/y_b)\tan\beta$.  That is, in parameter regimes where $\tan\beta\gg 1$, the impact of the $\Delta_b$ correction can be significantly enhanced, thereby modifying the couplings of the neutral Higgs bosons to $b$-quark pairs~\cite{Carena:2002es}.

In obtaining Eqs.~(\ref{susy1:eq:eqmH1up0}) and (\ref{Deltab}), the $\mu$ parameter and the potentially complex SUSY-breaking parameters were assumed to be real.
In the so-called complex MSSM, where $A_t$, $A_b$, $\mu$, and the gaugino mass parameters can be complex, there generically exist unremovable complex phases that provide new sources of CP violation.
Thus, the radiatively-corrected Higgs sector, which now depends on these new CP-violating phases,
is no longer CP conserving (\eg, the neutral Higgs scalars are no longer CP-eigenstates). Further details on the complex MSSM Higgs sector can be found in Ref.~\cite{Pilaftsis:1999qt,*Carena:2000yi,*Choi:2000wz, *Carena:2000ks,*Heinemeyer:2001qd,*Carena:2001fw,*Frank:2006yh,*Heinemeyer:2007aq}. 

In the NMSSM with $m_h\simeq 125$~GeV, the dominant radiative correction
to \Eq{susy1:eq:nmssmmass} 
is the
same as the one given in \Eq{susy1:eq:eqmH1up0}.  However, in contrast to the MSSM, one
does not need as large a boost from radiative corrections to achieve
a Higgs mass of $125$~GeV in certain regimes of the NMSSM parameter
space (\eg, $\tan\beta\sim 2$ and $\lambda\sim 0.7$\cite{Carena:2015moc}{}).

\subsection{The hMSSM approximation}\index{hMSSM}
\label{hMSSM}

As exhibited in Ref.~\cite{Haber:1996fp}, the dominant one-loop radiative correction to the value of $m^2_h$ in \Eq{susy1:eq:eqmH1up0} that yields the term proportional to $m_t^4\ln(M_S^2/m_t^2)$ is due entirely to the radiatively-corrected
$22$ element of the CP-even Higgs squared-mass matrix $\mathcal{M}^2_{22}$.  Thus, 
the authors of Refs.~\cite{Djouadi:2013uqa,Djouadi:2015jea} 
suggested a simple recipe to account for the leading radiative corrections to the neutral CP-even Higgs sector of the MSSM without explicitly fixing the parameters of the top squark sector.  In this recipe,
the expressions for $\mathcal{M}^2_{11}$ and $\mathcal{M}^2_{12}$ given in \Eq{CPevenSqMass} are retained, whereas $\mathcal{M}^2_{22}$ is left unspecified.   After diagonalizing the resulting squared-mass matrix, expressions of the form are obtained
\begin{align}
m_{h,H}^2 &= f_\pm(m_A^2,m_Z^2,\tan\beta,\mathcal{M}_{22}^2)\,,\label{hmssm1} \\
\cos\alpha &= g(m_A^2,m_Z^2,\tan\beta,\mathcal{M}_{22}^2)\,,\label{hmssm2}
\end{align}
where the functions $f_\pm$ and $g$ are the result of the diagonalization procedure described above.  One can use \Eq{hmssm1} to solve
for $\mathcal{M}^2_{22}$ in terms of the experimentally measured Higgs mass, $m_h\simeq 125$~GeV.  One then uses Eqs.~(\ref{hmssm1}) and (\ref{hmssm2})
to obtain expressions for $m_H^2$ and $\cos\alpha$ that are functions of the \textit{measured} mass $m_h$, which replace the tree-level results given in Eqs.~(\ref{susy1:eq:treelevelHmass}) and (\ref{susy1:eq:treelevelangle}).
This framework was dubbed the hMSSM in \Ref{Djouadi:2013uqa}.  
\index{hMSSM}%
Having fixed $m_h\simeq 125$~GeV, the Higgs phenomenology of the hMSSM is entirely governed by two MSSM input parameters, $m_A$ and $\tan\beta$. For example, one must assume that the radiative
corrections to the Yukawa Lagrangian [e.g., the $\Delta_b$ correction exhibited in \Eq{Deltab}] are negligible.

Although the hMSSM can be readily applied to LHC data to constrain the
neutral CP-even Higgs sector of the MSSM, it can lead to results that are not robust in
a more general MSSM parameter scan. 
Indeed, a more complete treatment of the radiative corrections can yield results that cannot
be accounted for in the hMSSM framework.  
Examples of benchmark points in the MSSM parameter space that cannot be reproduced by
the hMSSM analysis are examined in Refs.~\cite{Lee:2015uza,Haber:2017erd,Bahl:2019ago}.  Finally, we note that the hMSSM framework is not relevant for describing the shift in the properties of the charged Higgs boson due to radiative corrections.\index{hMSSM}

\subsection{The Higgs alignment limit and SUSY}\index{Higgs alignment limit}\index{alignment limit}
\label{alignment}

In the Higgs alignment limit of an extended Higgs sector~\cite{Gunion:2002zf,Craig:2012vn,*Craig:2013hca,*Carena:2013ooa,*Haber:2013mia,*BhupalDev:2014bir}, the tree-level properties of one of the neutral scalar states match those of the SM Higgs boson.  In light of the LHC Higgs data (see Ref.~\cite{PDGhiggs}), where the Higgs boson is observed to be SM-like~\cite{ATLAS:2022vkf,CMS:2022dwd}, 
any supersymmetric extension of the SM must account for a SM-like Higgs boson. 

Starting from \Eq{susy1:eq:Hpot},
we can define two linear combinations of the MSSM Higgs fields $H_u$ and $H_d$:
\begin{equation}
    \mathcal{H}_1\equiv H_d\cos\beta+H_u\sin\beta\,,
    \qquad\quad
    \mathcal{H}_2 \equiv -H_d\sin\beta+H_u\cos\beta\,.
\end{equation}
Note that the corresponding VEVs of these fields are $\langle \mathcal{H}_1^0\rangle=v/\sqrt{2}$ and $\langle \mathcal{H}_2^0\rangle=0$, respectively.  If the neutral scalar field defined by $\varphi\equiv \sqrt{2}\,{\rm Re}\,\mathcal{H}_1^0-v$ were an eigenstate of the CP-even neutral scalar squared-mass matrix
(corresponding to the alignment in field space of $\varphi$ with the direction of the VEV), then
it would follow that $\cos(\beta-\alpha)=0$, under the assumption that the observed Higgs boson is the lighter of the two CP-even scalars~\cite{Bernon:2015qea}.  One can check that in the alignment limit described above,
the tree-level properties of $\varphi$ coincide with those of the SM Higgs boson.  

To achieve the Higgs alignment limit, one must suppress the mixing of the field $\varphi$ and the second orthogonal CP-even scalar field.  This can be achieved in two different scenarios.
In the so-called decoupling limit of the extended Higgs sector, one neutral scalar mass eigenstate is identified with the SM-like Higgs boson, and all other Higgs scalar mass eigenstates are assumed to be much heavier.   As previously noted below \Eq{cbma}, $m_H\gg m_h$ yields $|\cos(\beta-\alpha)|\ll 1$, which is consistent with current LHC experimental bounds~\cite{ATLAS:2024lyh}.  If the decoupling limit is realized in the MSSM, then $H$, $A$ and $H^\pm$ must be significantly heavier (most likely with masses of order 500 GeV or larger~\cite{ATLAS:2024lyh}).
Indeed, Higgs alignment via decoupling is generic and can be achieved in many extensions of the SM.

The Higgs alignment limit can also be realized without decoupling if the value of the off-diagonal element of the CP-even squared-mass matrix (expressed with respect to the $\mathcal{H}_1$--$\mathcal{H}_2$ field basis)
is much smaller than the corresponding diagonal elements.  This cannot be achieved at tree-level in a viable region of the MSSM parameter space (nor in the hMSSM framework discussed in \Sec{hMSSM}). However, regions of the MSSM parameter space
do exist, albeit quite fine-tuned,
in which Higgs alignment without decoupling is achieved once radiative corrections are taken into account~\cite{Haber:2017erd}.  The NMSSM provides a more robust scenario for Higgs alignment without decoupling as shown in Ref.~\cite{Carena:2015moc}, where
the discovery of additional Higgs scalar states at future runs at the LHC would be expected. 

As recently reiterated in Ref.~\cite{Constantin:2025mex}, the precise measurements of the Higgs boson properties, which would be sensitive to small departures from the Higgs alignment limit, offer a valuable and complementary probe of weak-scale SUSY models, 
due to the radiative effects of superpartners  on Higgs couplings, even in scenarios where direct searches for superpartners
are not effective.

\section{Restricting the MSSM parameter freedom}\label{susy1:sec:restrict-MSSM}

In Sections~\ref{susy1:sec:sp-spectrum} and~\ref{susy1:sec:susy-higgs}, we surveyed the parameters that comprise
the MSSM-124.  However, the MSSM-124 is not a phenomenologically viable theory over much of its parameter space.
In particular, a generic point of
the MSSM-124 parameter space exhibits:
(i)~no conservation of the separate lepton numbers
$L_e$, $L_\mu$, and $L_\tau$; (ii)~unsuppressed
flavor-changing neutral currents (FCNCs);
and (iii)~new sources of CP~violation\cite{Khalil:2002qp} that are
inconsistent with the experimental bounds.

In addition, one-loop radiative corrections can introduce
CP-violating effects in the Higgs sector that depend on
some of the CP-violating phases among the MSSM-124
parameters\cite{Demir:1999hj}{}.
This phenomenon is most easily understood in a scenario where $m_A\ll M_S$ 
(i.e., all five physical Higgs states are significantly lighter than the SUSY 
breaking scale).  In this case, one can integrate out the heavy superpartners 
to obtain a low-energy effective theory with two Higgs doublets.  The resulting 
effective two-Higgs doublet model will now contain all possible Higgs self-interaction 
terms (both CP-conserving and CP-violating) and Higgs-fermion interactions (beyond 
those of Type-II) that are consistent with electroweak gauge invariance\cite{Haber:2007dj}.

As noted above, the
MSSM contains new sources of CP violation.  Indeed,
for TeV-scale sfermion and gaugino masses,
some combinations of the complex phases of the gaugino-mass
parameters, the $A$-parameters, and $\mu$ must be less than about
$10^{-2}$--$10^{-3}$ to avoid generating electric
dipole moments for the neutron, electron, and atoms~\cite{Fischler:1992ha,*susy1:MasieroCP,*Pospelov:2005pr}{}
in conflict with observed
data\cite{Abel:2020pzs,*Roussy:2022cmp}.
The rarity of FCNCs\cite{Gabbiani:1996hi,*RamseyMusolf:2006vr,Carena:2008ue} places
additional constraints on the off-diagonal matrix elements of
the squark and slepton soft-SUSY-breaking squared-masses
and $A$-parameters (see \Sec{susy1:sec:MSSM-124}).

The MSSM-124 is also theoretically incomplete
as it provides no explanation for the fundamental origin of the
super\-symmetry-breaking parameters.
The successful unification of the MSSM gauge couplings\index{GUT (grand unified theory)}
at the GUT scale, $M_{\rm GUT}\sim 10^{16}$~GeV, close to the Planck
scale\cite{susy1:Polonsky,ArkaniHamed:2004fb,*Giudice:2004tc,susy1:Mohapatra03,susy1:raby},
\begin{equation}
g_s(M_{\rm GUT}) = g(M_{\rm GUT})=\sqrt{\frac{5}{3}}\, g^\prime(M_{\rm GUT}), \label{gutU}
\end{equation}
suggests that the high-energy structure of the theory
may be considerably simpler than its low-energy
realization.\footnote{Generically, the normalization of the U(1) hypercharges exhibited in Table~\ref{susy1:tab:MSSM} is a matter of convention.  In particular, the U(1) hypercharges can be rescaled by absorbing the scaling factor into a redefinition of the hypercharge gauge coupling $g^\prime$.  However in a grand unified theory (GUT), the embedding of the hypercharge U(1) generator into the Lie algebra of a (unified) simple gauge group fixes the normalization of the U(1) hypercharges and results in the rescaled hypercharge gauge coupling shown in \Eq{gutU}.}
In a top-down approach, the dynamics that governs the theory at high energies
is used to derive the effective broken-supersymmetric
theory at the TeV scale. 

In this Section, we examine a number
of theoretical frameworks that potentially yield
phenomenologically viable regions of
the MSSM-124 parameter space.  The resulting
supersymmetric particle spectrum is then a function of a relatively
small number of input parameters.  This is accomplished by imposing a
simple structure on the soft SUSY-breaking  parameters at
a common high-energy scale $M_X$ (typically chosen to be
the Planck scale, $M_P$, the GUT scale, $M_{\rm GUT}$,
or the messenger scale, $M_{\rm mess}$).  These serve as initial
\index{messenger scale}%
conditions for the MSSM
\index{RGE, renormalization group equation}%
\index{Renormalization group equation (RGE)}%
renormalization group equations (RGEs), which
are given in the two-loop approximation
in \Ref{Martin:1993zk,*Fonseca:2011vn,*Staub:2010jh}. An automated program to
compute RGEs for the MSSM and other supersymmetric models of
new physics has been developed in \Ref{Staub:2013tta,*Staub:2015kfa,*susy1:sarah-sr2,*Fonseca:2011sy,*susy1:susyno}.
Solving these equations numerically,  one can then
derive the low-energy MSSM parameters relevant for phenomenology.
A number of software packages exist that numerically calculate the
spectrum of supersymmetric particles,
consistent with theoretical conditions on SUSY breaking at
high energies and some experimental data at low energies\cite{Allanach:2001kg,*susy1:softsusy-sr1,*Djouadi:2002ze,*susy1:suspect-sr1,*Paige:2003mg,*susy1:isajet-sr1,*Porod:2003um,*susy1:spheno-sr1,*Athron:2014yba,*susy1:flexiblesusy-sr1,Slavich:2020zjv}.

Examples of viable frameworks are provided by
models of gravity-mediated,
anomaly-mediated, and gauge-mediated
SUSY breaking.
In some of these approaches, one of the diagonal Higgs
squared-mass
parameters is driven negative by renormalization group
evolution\cite{Ibanez:1982fr}{}.  In such models, electroweak symmetry breaking is generated
radiatively, and the resulting electroweak symmetry-breaking scale is
intimately tied to the scale of low-energy SUSY breaking.

\subsection{Gaugino mass relations}\label{susy1:sec:gaugino-mass}
\index{gaugino mass unification}%
One prediction of many supersymmetric grand unified models
is the unification of the (tree-level)
gaugino mass parameters\footnote{Non-universal gaugino mass
parameters can also be a viable option in grand unified models with
non-minimal gauge kinetic functions~\cite{Drees:1985bx,*Ellis:158128,*Martin:2009ad}.} at some high-energy scale, $M_X$,
\begin{equation}\label{susy1:eq:gunif}
M_1(M_X) = M_2(M_X) = M_3(M_X) = m_{1/2}\,.
\end{equation}
Due to renormalization group running, in the one-loop approximation the effective low-energy gaugino mass
parameters (at the electroweak scale) are related,
\begin{equation}\label{susy1:eq:eqmass3}
M_3=(g_s^2/g^2)M_2\simeq 3.5M_2\ ,\quad\,\,\,
M_1=(5g^{\prime\,2}/3g^2)M_2\simeq 0.5M_2.
\end{equation}
\Eq{susy1:eq:eqmass3} can arise more generally in gauge-mediated
SUSY-breaking models where the gaugino masses are generated
at the messenger scale $M_{\rm mess}$ (which typically lies
significantly below the unification scale where the gauge couplings
unify).  In this case, the gaugino mass parameters
are proportional to the corresponding
squared gauge couplings at the messenger scale.

When \Eq{susy1:eq:eqmass3} is satisfied, the chargino and neutralino masses and
mixing angles depend only on three unknown parameters: the gluino mass,
$\mu$, and $\tan\beta$.  It then follows that the lightest neutralino
must be heavier than 46 GeV due to the non-observation of
charginos at LEP\cite{Abdallah:2003xe}{}.
If in addition $|\mu|\gg |M_1|\gsim\mZ$, then the
lightest neutralino is nearly a pure bino, an assumption often made
in supersymmetric particle searches at colliders.
Although \Eq{susy1:eq:eqmass3} is often assumed in many
phenomenological studies, a truly model-independent approach
would take the gaugino mass parameters $M_1$, $M_2$, and $M_3$ to be independent
parameters to be determined by experiment.  Indeed, an
approximately massless neutralino
{\it cannot}\/ be ruled out at present by a model-independent
analysis\cite{Dreiner:2009ic}.

It is possible that the tree-level masses of the gauginos are zero.
In this case, the gaugino mass parameters arise at one-loop and do not
satisfy \Eq{susy1:eq:eqmass3}.  For example,
the gaugino masses in AMSB models arise entirely from a model-independent
contribution derived from the super-conformal
anomaly\cite{Randall:1998uk,Giudice:1998xp,*Pomarol:1999ie,*Jung:2009dg}{}.  In this case,
\Eq{susy1:eq:eqmass3} is replaced (in the one-loop approximation) by:
\begin{equation}\label{susy1:eq:eqanom}
M_i\simeq {\frac{b_i g_i^2}{16\pi^2}m_{3/2}}\,,
\end{equation}
where $m_{3/2}$ is the gravitino mass
and the $b_i$ are the coefficients of the MSSM gauge beta-functions
corresponding to the corresponding U(1), SU(2), and SU(3) gauge groups,
$(b_1,b_2,b_3)= (\frac{33}{5},1,-3)$.
\Eq{susy1:eq:eqanom} yields
$M_1\simeq 2.8M_2$ and $M_3\simeq -8.3M_2$, which implies that
the lightest chargino pair and neutralino comprise a
nearly mass-degenerate triplet of winos, $\widetilde W^\pm$,
$\widetilde W^0$ (cf.~Table 1),
over most of the MSSM parameter space.  For example, if $|\mu|\gg
m_Z$, $|M_2|$, then \Eq{susy1:eq:eqanom} implies that $M_{\chinopm}\simeq M_{\ninoone}\simeq
M_2$\cite{Gunion:1987yh,*Choi:2004rf}.
Alternatively, one can construct an AMSB model where
$|\mu|, m_Z\ll M_2$, which yields an LSP that is an approximate higgsino
state\cite{Baer:2018hwa}.  In both cases, the
corresponding supersymmetric phenomenology differs significantly
from the standard phenomenology
based on \Eq{susy1:eq:eqmass3}\cite{Feng:1999fu,*Gunion:1999jr,Gherghetta:1999sw}.

Finally, it should be noted that the
unification of gaugino masses (and scalar masses) can be accidental.
In particular, the energy scale where unification takes place may not be
directly related to any physical scale.  One version of this phenomenon has been
called mirage unification and can occur in certain theories
of fundamental SUSY breaking\cite{Endo:2005uy,*Choi:2005uz,*LoaizaBrito:2005fa}.

\subsection{Constrained versions of the MSSM: mSUGRA, CMSSM, etc.}\label{susy1:sec:mSUGRA-etc}

\index{mSUGRA, minimal supergravity}%
\index{Minimal supergravity, mSUGRA}

In the minimal supergravity (mSUGRA)
framework\cite{Nilles:1983ge,susy1:Weinberg00,susy1:Nath,susy1:DHM,susy1:sugramodels,Alvarez-Gaume:1983drc,Hall:1983iz}, 
the minimal form of the K\"ahler potential
is employed, which yields standard kinetic energy terms for
the MSSM fields\cite{susy1:bim}.  As a result, the
soft super\-symmetry-breaking parameters at the high-energy scale
$M_X$ take
a particularly simple form in which the scalar squared-masses
and the $A$-parameters are flavor-diagonal and universal\cite{Hall:1983iz}:
\begin{equation}\label{susy1:eq:plancksqmasses}
\begin{aligned}
 &M^2_{\widetilde{Q}} (M_X) = M^2_{\widetilde{U}}(M_X) = M^2_{\widetilde{D}}(M_X) =  m_0^2
            {\bf 1}\,,\\
 &M^2_{\widetilde{L}}(M_X) = M^2_{\widetilde{E}}(M_X) = m_0^2 {\bf 1} \,,\\
 &m^2_{H_u}(M_X) = m^2_{H_d}(M_X) = m_0^2 \,,\\
 & A_U(M_X) = A_D(M_X) = A_E(M_X) = A_0 {\bf 1}\,,
\end{aligned}
\end{equation}
where ${\bf 1}$ is a $3\times 3$ identity matrix in generation space.
As in the SM, this approach exhibits minimal flavor violation 
(\emph{e.g.}, see \Ref{DAmbrosio:2002vsn,*Smith:2009hj}{}),
whose unique source is
the nontrivial flavor structure of the Higgs-fermion Yukawa couplings.
The gaugino masses are also unified according to \Eq{susy1:eq:gunif}.

Renormalization group evolution is then used to derive the values of the
supersymmetric parameters at the low-energy (electroweak) scale.  For
example, to compute squark masses, one should use the
low-energy values for $M_{\widetilde{Q}}^2$, $M_{\widetilde{U}}^2$,
and $M_{\widetilde{D}}^2$ in \Eq{susy1:eq:squarkmatrix}. Through
the renormalization group running with boundary conditions specified in
\Eq{susy1:eq:eqmass3} and \Eq{susy1:eq:plancksqmasses}, one can show that
the low-energy values of $M_{\widetilde{Q}}^2$, $M_{\widetilde{U}}^2$,  and
$M_{\widetilde{D}}^2$ depend primarily on $m_0^2$ and $m_{1/2}^2$.
A number of useful
approximate analytic expressions for superpartner masses in terms
of the mSUGRA parameters can be found in \Ref{susy1:Drees95}.

One can count the number of independent parameters in the
mSUGRA framework.  In addition to 18 SM parameters
(excluding the Higgs mass), one must specify
$m_0$, $m_{1/2}$, $A_0$, the Planck-scale values for $\mu$ and
$B$-parameters (denoted by $\mu_0$ and $B_0$), and
the gravitino mass $m_{3/2}$.  Without additional model assumptions,
$m_{3/2}$ is independent of the parameters that govern the
mass spectrum of the superpartners of the SM\cite{Hall:1983iz}{}.
In principle,
$A_0$, $B_0$, $\mu_0$, and $m_{3/2}$ can be complex, although in
the mSUGRA approach, these parameters are taken 
to be real for simplicity.

As previously noted,
renormalization group evolution is used to compute the low-energy
values of the mSUGRA parameters, which then fixes all the
parameters of the low-energy MSSM\@.  In particular, the two Higgs
VEVs (or equivalently, $m_Z$ and $\tan\beta$)
can be expressed as a function of the Planck-scale supergravity
parameters.
In light of \Eq{susy1:eq:minbeta} and \Eq{susy1:eq:minconditions},
a common procedure
is
to determine $\mu_0$ and $B_0$ in terms of $m_Z$ and $\tan\beta$
[the
sign of $\mu_0$, denoted ${\rm sgn}(\mu_0)$ below, is not fixed in this
process].  In this case, the MSSM spectrum
and its interaction strengths are fixed by five parameters:
\begin{equation}\label{susy1:eq:sugraparms}
m_0\,,\, A_0\,,\, m_{1/2}\,,\, \tan\beta\,,\, {\rm and}~{\rm
sgn}(\mu_0)\,,
\end{equation}
and an independent gravitino mass $m_{3/2}$ (in
addition to the 18 parameters of the SM).
In \Ref{Kane:1993td}, this framework was dubbed the
constrained minimal supersymmetric extension of the SM (CMSSM).
\index{CMSSM, Constrained MSSM}%
\index{Constrained MSSM, CMSSM}%
Additional relations such as $B_0=A_0-m_0$ and $m_{3/2}=m_0$ comprise the
original mSUGRA proposal\cite{susy1:sugramodels,susy1:bim,Ellis:2003pz,*Ellis:2004qe}.

One can also relax the universality of scalar masses by
decoupling the squared-masses of the Higgs bosons and the squarks/sleptons.
This leads to the non-universal Higgs mass models (NUHMs), thereby adding
one or two new parameters to the CMSSM depending on whether the
diagonal Higgs scalar squared-mass parameters ($m_{H_d}^2$ and $m_{H_u}^2$)
are set equal (NUHM1~\cite{Baer:2004fu}) or taken to be independent (NUHM2~\cite{Berezinsky:1995cj,*Ellis:2002iu}) at the
high energy scale $M_X$.  Clearly, this modification preserves
the minimal flavor violation of the mSUGRA approach.
Nevertheless, the mSUGRA approach and its NUHM generalizations are probably too simplistic.
Theoretical considerations suggest that the universality of Planck-scale
soft SUSY-breaking parameters is not generic\cite{Ibanez:1992hc,*deCarlos:1992pd,*Kaplunovsky:1993rd,*Brignole:1993dj}{}.
In particular, effective operators at the Planck scale exist
that do not respect flavor universality, and it is difficult to
find a theoretical principle that would forbid them.

In the framework of supergravity, if anomaly mediation is the sole source
of SUSY breaking, then the gaugino mass parameters, diagonal scalar
squared-mass parameters, and the SUSY-breaking
trilinear scalar interaction terms (proportional to $\lambda_f A_F$)
are determined in terms of the beta functions of the gauge and Yukawa
couplings and the anomalous dimensions of the squark and slepton
fields\cite{Randall:1998uk,Giudice:1998xp,*Pomarol:1999ie,*Jung:2009dg,Gherghetta:1999sw}{}.
As noted in \Sec{susy1:sec:hidden-sect}, this approach yields tachyonic sleptons
in the MSSM unless additional sources of
SUSY breaking are present.  In the
minimal AMSB (mAMSB) 
\index{mAMSB, minimal AMSB}%
\index{minimal AMSB, mAMSB}%
scenario,
a universal squared-mass parameter, $m_0^2$, is added to the
AMSB expressions for the diagonal scalar
squared-masses\cite{Gherghetta:1999sw}{}.  Thus, the mAMSB spectrum and its
interaction strengths are determined by four parameters, $m_0^2$, $m_{3/2}$, $\tan\beta$ and ${\rm sgn}(\mu_0)$.

The mAMSB scenario appears to be ruled out based on
the observed value of the Higgs boson mass, assuming an upper limit on
$M_S$ of a few TeV, since the mAMSB constraint on $A_F$
implies that the maximal mixing scenario cannot be achieved [cf.~\Eq{susy1:eq:eqmH1up0}].  Indeed,
under the stated assumptions, the mAMSB Higgs mass upper bound lies
below the observed Higgs mass value\cite{Arbey:2013kgl}.
Thus within the mAMSB scenario, either an
additional SUSY-breaking contribution to $\lambda_f A_F$, and/or
new ingredients beyond the MSSM are required.  

\subsection{Gauge-mediated SUSY breaking}\label{susy1:sec:gauge-med-sb}

\index{GMSB} \index{gauge-mediated supersymmetry breaking}%
In contrast to
models of gravity-mediated SUSY breaking, the flavor universality of
the fundamental soft SUSY-breaking squark and slepton
squared-mass parameters is guaranteed in gauge-mediated supersymmetry breaking (GMSB) because the super\-symmetry breaking is communicated
to the sector of MSSM fields via gauge
interactions\cite{Dine:1993yw,Giudice:1998bp}{}.  In GMSB models, the
mass scale of the messenger sector (or its equivalent) is
sufficiently below the Planck scale such
that the additional SUSY-breaking effects mediated by
supergravity can be neglected.  

In the minimal GMSB approach,
there is one effective mass scale, $\Lambda$, that determines all
low-energy scalar and gaugino mass parameters through loop effects,
while the resulting $A$-parameters are suppressed.
In addition, the minimal form of the K\"{a}hler potential is employed.
Assuming that the superpartner masses
are no larger than a few TeV, one must take $\Lambda\sim {\mathcal O}(100$~TeV$)$.
The origin of the $\mu$ and $B$-parameters is model-dependent, and
lies somewhat outside the purview
of gauge-mediated SUSY breaking\cite{Dvali:1996cu}.

The simplest GMSB models appear to be ruled out based on
the observed value of the Higgs boson mass.  Due to
suppressed $A$-parameters, it is difficult to boost
the contributions of the radiative corrections in \Eq{susy1:eq:eqmH1up0} to obtain
a Higgs mass as large as 125 GeV, under the assumption that $M_S$ is no larger than a few TeV.  However, this conflict
can be alleviated in more complicated GMSB models\cite{Draper:2011aa}{}.
To analyze these generalized GMSB models, it has been especially
fruitful to develop model-independent techniques that encompass all known
GMSB models\cite{Meade:2008wd,*Buican:2008ws}.
These techniques are well-suited for a comprehensive analysis\cite{Rajaraman:2009ga,*Carpenter:2008wi}
of the phenomenological profile of gauge-mediated SUSY breaking.

The gravitino is the LSP in minimal GMSB models, as noted
in \Sec{susy1:sec:hidden-sect}. 
As a result, the next-to-lightest
supersymmetric particle (NLSP)
\index{NLSP, next-to-lightest supersymmetric particle}%
\index{next-to-lightest supersymmetric particle, NLSP}%
now plays a crucial role in the phenomenology
of supersymmetric particle production and decays.
Note that unlike the LSP, the NLSP can be charged.
In GMSB models,
the most likely candidates for
the NLSP are $\ninoone$ and $\widetilde\tau_R^\pm$.  The NLSP
will decay into its superpartner plus a gravitino
(\eg, $\ninoone\to\gamma\widetilde G$,
$\ninoone\to Z\widetilde G$, $\ninoone\to h^0\widetilde G$ or
$\widetilde\tau_{{1}}^\pm\to\tau^\pm\widetilde G$), with lifetimes
and branching ratios that depend on the model parameters.
There are also GMSB scenarios in which there are several nearly degenerate
co-NLSPs, any one of which can
be produced at the penultimate step of a supersymmetric decay
chain\cite{Ambrosanio:1997bq}{}.  For example, in the slepton co-NLSP case, all three
right-handed sleptons are close enough in mass and thus can each play the role
of the NLSP.

Different
choices for the identity of the NLSP and its decay rate lead to a
variety of distinctive supersymmetric
phenomenologies\cite{Giudice:1998bp,susy1:gmsb}{}.  For
example, a long-lived
$\ninoone$-NLSP that decays outside collider detectors leads to
supersymmetric decay chains with missing energy in association with
leptons and/or hadronic jets (this case is indistinguishable from the
standard phenomenology of the $\ninoone$-LSP).  On the other hand, if
$\ninoone\to\gamma\widetilde G$ is the dominant decay mode, and the
decay occurs inside the detector, then nearly {\it all}\/ supersymmetric
particle decay chains would produce a photon.  In contrast, in the
case of a $\widetilde\tau_{{1}}^\pm$-NLSP, the $\widetilde\tau_{{1}}^\pm$ would either be
long-lived or would decay inside the detector into a $\tau$-lepton
plus missing energy.

A number of attempts have been made to address the origins of the $\mu$ and $B$-parameters
in GMSB models based on the field content
of the MSSM ({\it e.g.}, see Refs.~\cite{Dvali:1996cu,Roy:2007nz}).
An alternative approach is to consider GMSB models
based on the NMSSM\cite{deGouvea:1997cx}{}.  The VEV of the additional singlet Higgs superfield can be used
to generate effective $\mu$ and $B$-parameters\cite{Han:1999jc,*Chacko:2001km,*Delgado:2007rz,*Liu:2008pa}{}.   Such models provide an alternative GMSB framework for achieving a Higgs mass of 125 GeV, while still being consistent with LHC bounds on supersymmetric particle masses.

\subsection{The phenomenological MSSM}\label{susy1:sec:phenom-MSSM}

\index{pMSSM, phenomenological MSSM}%
\index{Phenomenological MSSM, pMSSM}%
Any of the theoretical assumptions described in the previous three subsections
must be tested experimentally and could turn out to be wrong.
To facilitate the exploration of MSSM phenomena in a more
model-independent way while respecting the constraints noted at the
beginning of \Sec{susy1:sec:restrict-MSSM}, the phenomenological MSSM (pMSSM) has
been introduced\cite{susy1:pmssm,*Berger:2008cq,*Allanach:2015cia}.

The pMSSM is governed by 19 independent real supersymmetric
parameters: the three gaugino
mass parameters $M_1$, $M_2$ and $M_3$, the Higgs sector parameters $m_A$ and
$\tan\beta$, the Higgsino mass parameter $\mu$, five sfermion
squared-mass parameters for the degenerate first and second
generations ($M^2_{\widetilde Q}$, $M^2_{\widetilde U}$,  $M^2_{\widetilde D}$,
$M^2_{\widetilde L}$ and $M^2_{\widetilde E}$), the five
corresponding sfermion squared-mass parameters for
the third generation, and three third-generation $A$-parameters
($A_t$, $A_b$ and $A_\tau$).
The
first and second generation $A$-parameters are typically
neglected in pMSSM studies, as their
phenomenological consequences are negligible in most applications.
One counterexample arises when considering the $A_\mu$ dependence of
the anomalous magnetic moment of the muon, which can be as significant
as other contributions due to superpartner mediated radiative 
corrections~\cite{Martin:2001st}.
Consequently, the original pMSSM
approach has been extended
to include a 20th parameter,
$A_\mu$~\cite{AbdusSalam:2009qd}.
Other pMSSM extensions that include CP-violating
SUSY-breaking parameters have been considered in \Ref{Berger:2015eba}.

The 19-parameter pMSSM is often further constrained to expedite scans over the
parameter space.
For example, in \Ref{deVries:2015hva}, the number of pMSSM
parameters is reduced to ten by assuming one common squark squared-mass parameter
for the first two generations, a second common squark squared-mass parameter
for the third generation, a common (charged)
slepton squared-mass parameter and a common
third generation $A$parameter.
In \Ref{Bagnaschi:2017tru}
an eleven parameter pMSSM is defined by allowing for a different
stau squared-mass parameter from that of the first two generation charged sleptons.
Other applications of the pMSSM approach (with a reduced pMSSM parameter space) to 
supersymmetric particle searches, 
and a discussion of the implications for past and future LHC and dark
matter studies can be found in Refs.~\cite{deVries:2015hva,Cahill-Rowley:2014wba,*Cahill-Rowley:2014boa,*Bertone:2015tza}.

\subsection{Simplified models}\label{susy1:sec:simplified-models}

\index{simplified models}%
As discussed in Ref.~\cite{SUSYpdgEXP}, the
experimental collaborations present the results of their searches for
supersymmetric particles primarily in terms of simplified models.
Simplified models for supersymmetric particle searches\cite{ArkaniHamed:2007fw,*Alwall:2008va,*Alwall:2008ag,*Alves:2010za,*Alves:2011sq,*Alves:2011wf}
are defined mostly by the empirical objects
and kinematic variables involved in the search.
The interpretation of the experimental results
usually involves only a small number of
supersymmetric particles (often two or three).
Other supersymmetric particles are
assumed to play no role (this may happen by virtue of them being too heavy
to be produced). Experimental bounds from the non-observation of a signal are
usually presented in terms of the physical masses of the supersymmetric
particles involved.
Bounds on the relevant supersymmetric particle masses
may be presented assuming
values for the branching ratio of certain supersymmetric particle decays, or
as an upper bound on the 
signal production cross section as a function of the relevant
supersymmetric particle masses. 

For example, consider a search for hadronic jets plus missing transverse momentum.
One can match such a search to a simplified model of squark pair
production followed by the subsequent decay of each 
squark into a quark (which appears as a jet) and a neutralino LSP
that produces the missing transverse momentum,
{\it i.e.}
$\squark \squark \rightarrow (q \ninoone) (q \ninoone)$.
Excluded cross sections resulting from the non-observation of a  
signal (which in this case could consist of some specified minimum value of missing
transverse momentum and at least two hard jets)
may be exhibited in the
squark mass versus LSP mass plane. 

The large number of free parameters that govern a typical supersymmetric model 
makes it difficult to present experimentally excluded regions in any generality.  This is where
simplified models have an apparent advantage in
that they depend on far fewer free parameters than
more complete supersymmetric models. 
However, if limits are quoted on supersymmetric particle masses without reference to
the signal production cross section from a simplified model analysis, then there
are several potential pitfalls.  For example, chargino/neutralino mixing affect their production cross sections.
Moreover, mass limits can differ from those obtained in full models because 
there may be contributions
to the signal coming from processes involving 
supersymmetric particles other than those assumed. 
Indeed, in the
$\squark \squark \rightarrow q \ninoone q \ninoone$ process mentioned above, 
the simplified model analysis does not account for the interference with
tree-level $t$-channel gluino contributions, nor does it account for other decay modes of the $\squark \squark$ pair. Nevertheless, 
simplified model bounds quoted purely in terms of supersymmetric particle masses
may still approximately hold over sizable regions of
parameter space of more complete models,
within which the simplified model is embedded. 
When simplified model limits are phrased as bounds on signal cross sections,
the aforementioned pitfalls are sidestepped. 

Simplified models thus remain an
efficient tool for organizing and presenting the results of supersymmetric 
particle searches.
A comparison between supersymmetric particle search
constraints in the context of simplified models and the corresponding
constraints obtained in the more complete pMSSM can be found
in~\Ref{Ambrogi:2017lov}. 

\section{Experimental data confronts the MSSM}\label{susy1:sec:exp-MSSM}

\index{MSSM current experimental status}%
At present, there is no significant 
evidence for weak-scale SUSY
from the data analyzed by the LHC experiments\cite{ATLAS:2024lda,*Sekmen:2025bxv}.
Recent LHC data
have been employed in ruling out the existence of colored
supersymmetric particles (primarily the gluino and the first
generation of squarks) with masses below about
2~TeV.
Moreover, given that the mass of the observed Higgs boson is 125 GeV, 
the results exhibited in Fig.~\ref{susy1:fig:kuts} tend to favor a mass scale of the top squarks somewhat above 2 TeV.
However, the precise mass limits are very model dependent. For example,
as Fig. 88.13 of Ref.~\cite{SUSYpdgEXP} demonstrates, 
regions of the pMSSM parameter space
can be identified in which lighter squarks and gluinos below 1~TeV
cannot be definitively ruled out.
Under the assumption of gaugino mass 
unification [cf.~\Eq{susy1:eq:eqmass3}], LHC searches result in a lower bound
on neutralino and chargino masses
of roughly 200~GeV.
It is also difficult to place general bounds on neutralino and chargino masses,
since the limits in terms of masses from direct searches tend to be
particularly model-dependent. 
It is therefore premature to rule out the entire 
framework of weak-scale supersymmetry
based on current LHC searches for supersymmetric particles~\cite{Constantin:2025mex}.  Nevertheless,
one must confront the tension that exists between the theoretical
expectations for the magnitude of the SUSY-breaking parameters
and the non-observation of supersymmetric
phenomena at colliders. 

\subsection{Naturalness constraints and the little hierarchy}\label{susy1:sec:natural-constr}

\index{naturalness}%
\index{little hierarchy}%
In \Sec{susy1:sec:Intro},
weak-scale SUSY was motivated as a natural solution
to the hierarchy problem, which could provide an understanding of
the origin of the electroweak symmetry-breaking scale without a significant
fine-tuning of the fundamental parameters that govern the MSSM\@.
\index{fine tuning}%
In this context, the weak-scale soft super\-symmetry-breaking masses
must be generally of the order of 1~TeV or below\cite{Barbieri:1987fn}{}.
This requirement is most easily seen in the determination of $m_Z$
by the scalar potential minimum condition.  In light of \Eq{susy1:eq:minconditions}, to avoid
the fine-tuning of MSSM parameters, the soft SUSY-breaking squared-masses
$m_{H_d}^2$ and $m_{H_u}^2$ and the higgsino squared-mass $|\mu|^2$ should all be roughly of ${\cal O}(m_Z^2)$.
Many authors have proposed quantitative measures of
fine-tuning\cite{Barbieri:1987fn,Ellis:1986yg,Anderson:1994dz,*Anderson:1994tr,*Anderson:1995cp,*Athron:2007ry,Cabrera:2008tj,*Baer:2012up,Feng:1999zg,Ghilencea:2012qk}.
One of the simplest measures is the one advocated by
Barbieri and Giudice\cite{Barbieri:1987fn}{} (which was also introduced previously
in \Ref{Ellis:1986yg}),
\begin{equation}\label{susy1:eq:finetune}
\Delta_i\equiv \left|{\frac{\partial\ln m_Z^2}{\partial\ln p_i}}\right|\,,\qquad
\Delta\equiv {\rm max}~\Delta_i\,,
\end{equation}
where the $p_i$ are the MSSM parameters at the high-energy scale
$M_X$, which are set by the fundamental SUSY-breaking dynamics.
The theory is more fine-tuned as $\Delta$ becomes larger.
However, different measures of fine-tuning yield quantitatively different 
results~\cite{Baer:2023cvi}; in particular, calculating minimal fine-tuning based on the
high-scale parameters [as defined in \Eq{susy1:eq:finetune}] yields a difference by a factor
of order 10 to fine-tuning based on TeV-scale
parameters~\cite{Baer:2013gva,vanBeekveld:2019tqp}. 

One can apply the fine-tuning measure to any explicit model of
SUSY breaking.  For example, in the approaches discussed
in \Sec{susy1:sec:restrict-MSSM}, the $p_i$ are
parameters of the model at the energy scale $M_X$ where
the soft SUSY-breaking operators are generated
by the dynamics of SUSY breaking.
Renormalization group evolution then determines the values of
the parameters appearing in \Eq{susy1:eq:minconditions} at the electroweak
scale.  In this way, $\Delta$ is sensitive to all the
SUSY-breaking parameters of the model
(see {\it e.g.}~\Ref{Kane:1998im,*BasteroGil:1999gu,*Casas:2003jx,*Abe:2007kf,*Essig:2007kh}).
The computation of $\Delta$ is often based on \Eq{susy1:eq:minconditions}, which is a tree-level condition.
However, the fine-tuning measure obtained at tree level can be somewhat reduced in value 
when loop corrections are included while remaining consistent with all experimental constraints~\cite{deCarlos:1993rbr,*Cassel:2009ps,*Cassel:2010px,*Ghilencea:2012gz,Ross:2017kjc}.

One way of taking fine-tuning into account
in fits to data using Bayesian statistics is to have a prior probability
distribution proportional to $1/\Delta$~\cite{Allanach:2006jc} so that
fine-tuning is balanced against the fit to empirical data. 
In such a Bayesian approach, it is important to choose the prior probability
distribution carefully, since prior probability densities that are flat in one
set of variables may not be flat in another, more fundamental set. One can in
fact derive a different measure of fine-tuning resulting from a Jacobian
factor when transforming to other variables.\footnote{For example, one may consider the parameters $\mu$ and
$m_{12}^2$ to be more fundamental than $\tan \beta$ and
$M_Z$. In this case, one would choose a flat prior probability distribution in
$\mu$ and $m_{12}^2$ rather 
than in $\tan \beta$ and $M_Z$~\cite{Allanach:2007qk,Cabrera:2008tj}. 
The Jacobian factor is then obtained from
Eqs.~(\ref{susy1:eq:minbeta})
and (\ref{susy1:eq:minconditions}).}
By comparing the results of 
several Bayesian fits with different (but reasonable) prior
probability distributions, one can assess the robustness of the fit with
respect to their variation, 
mitigating for subjectivity in the
interpretation of the fine-tuning measure.

As anticipated, there is a tension
between the present experimental lower limits on the masses of
colored supersymmetric particles\cite{Buchmueller:2013rsa,*Bechtle:2015nua,*Han:2016gvr,*GAMBIT:2017snp} and
the expectation that
super\-symmetry-breaking is associated with the electroweak
sym\-metry-breaking scale.
Moreover, in light of the results exhibited in \Fig{susy1:fig:kuts}, this tension is exacerbated by the observed value of the
Higgs mass ($m_h\simeq 125$~GeV),
which is suggestive of a value of $M_S$ in the multi-TeV range.  Indeed, if
$M_S$ characterizes the scale of all supersymmetric
particle masses, then one would crudely expect $\Delta\sim M_S^2/m_Z^2$.  For example, if $M_S\sim 2$~TeV then
one expects a fine-tuning of the MSSM parameters such that $\Delta^{-1}\sim 0.2\%$ 
to achieve the observed value of $m_Z$.
This separation of the electroweak symmetry-breaking and
SUSY-breaking scales is an example of the
little hierarchy problem\cite{Barbieri:2000gf,*Giusti:1998gz,*Cheng:2003ju,*Cheng:2004yc,*Harnik:2003rs}.

The fine-tuning parameter $\Delta$ can depend quite
sensitively on the structure of the SUSY-breaking
dynamics, such as the value of $M_X$ and relations among
SUSY-breaking parameters in the fundamental high
energy theory\cite{Baer:2012mv,*Baer:2012cf,*Feng:2013pwa}.
For example, in so-called focus point SUSY models\cite{Feng:1999mn,*Feng:2005hw,*Horton:2009ed,Feng:1999zg},
all squark masses can be as heavy as 5~TeV {\it without}\/ significant
fine-tuning.  This can be attributed to a focusing behavior of
the renormalization group evolution when
certain relations hold among the high-energy values of
the scalar squared-mass SUSY-breaking parameters.
Although the focus point region of the CMSSM still yields an uncomfortably
high value of $\Delta$ due to the observed Higgs
mass of 125 GeV, one can achieve moderate values of
$\Delta$ in models with NUHM2 boundary conditions for the scalar
masses\cite{Baer:2014ica}.

Among the colored superpartners,
the third generation squarks typically have the most significant
impact on the naturalness constraints\cite{Drees:1985jx,*Dimopoulos:1995mi,*Pomarol:1995xc}{}, while their masses
are the least constrained by the LHC data. Hence, in the absence
of any relation between third generation squarks and those of the
first two generations, the naturalness constraints due to present
LHC data can be considerably weaker than those obtained in the
CMSSM\@.
Indeed, models with first and second generation squark masses in the multi-TeV
range do not necessarily
 require significant fine tuning.  Such
models have the added benefit that undesirable FCNCs
mediated by squark exchange are naturally suppressed\cite{Dine:1990jd,*Cohen:1996vb}{}.
Other MSSM mass spectra that are compatible with moderate
fine tuning have been considered in 
Refs.~\cite{Baer:2012mv,Brust:2011tb}.

The lower bounds on squark and gluino masses may not be as large
as suggested by the experimental analyses based on the
CMSSM or simplified models. For example,
mass bounds for the gluino and the first and second generation squarks
based on the CMSSM
can often be evaded in alternative
or extended MSSM models, \eg, compressed SUSY\cite{Martin:2007gf,*Martin:2008aw}
and stealth SUSY\cite{Fan:2011yu,*Fan:2012jf}{}.
Moreover, the experimental lower limits
for the third generation squark masses
(which have a more direct impact on the fine-tuning measure)
are weaker than the corresponding mass limits for other colored
supersymmetric states.

Among the uncolored superpartners, the higgsinos are typically the most impacted by the naturalness constraints.
\Eq{susy1:eq:minconditions} suggests that the masses of the two neutral higgsinos
and charged higgsino pair (which are governed by $|\mu|$) should not be significantly larger than $m_Z$
to avoid an unnatural fine-tuning of the supersymmetric parameters, which
would imply the existence of light higgsinos (whose
masses are not well constrained, as they are
difficult to detect directly at the LHC due to their soft decay products).
However, it may be possible to avoid the conclusion that $\mu\sim
{\cal O}(m_Z)$ if additional
correlations among the SUSY breaking mass parameters and $\mu$ are present.
Such a scenario can be realized in models in which the boundary conditions for
SUSY breaking are generated by approximately
conformal strong dynamics.  For example, in the so-called scalar-sequestering model
of \Ref{Murayama:2007ge,*Perez:2008ng}, values of $|\mu|>1$~TeV can be achieved
while naturally maintaining the observed value of $m_Z$.

Finally, one can also consider extensions of the
MSSM in which the degree of fine-tuning is relaxed.
For example, it has already been noted in \Sec{susy1:sec:susy-higgs} that
it is possible to accommodate the observed Higgs mass
more easily in the NMSSM due to
contributions to $m_h^2$ proportional to the parameter $\lambda$.  This means
that one does not have to rely on a large contribution from 
radiative corrections to boost the Higgs mass sufficiently above its tree-level
bound.  This allows for smaller top squark masses, which are more
consistent with the demands of naturalness.  The reduction of the fine-tuning
in various NMSSM models was initially advocated in
\Ref{Dermisek:2005ar,*susy1:GunNMSSM-sr1,*Dermisek:2007yt}, and subsequently
treated in more detail in
Refs.~\cite{Hall:2011aa,Ross:2011xv,*Ross:2012nr,*Kaminska:2013mya}. 
Naturalness can also be relaxed in extended supersymmetric models with
vector-like quarks\cite{Martin:2012dg} and in gauge extensions of the MSSM\cite{Bellazzini:2009ix}.

The experimental absence of any new physics beyond the SM
at the LHC suggests that the principle of naturalness is presently under
significant stress~\cite{Dine:2015xga}.
Nevertheless, one must be very cautious when drawing conclusions
about the viability of weak-scale SUSY to explain the
origin of electroweak symmetry breaking, since different measures of
fine-tuning noted above can lead to different assessments~\cite{Baer:2023cvi,Baer:2013gva,vanBeekveld:2019tqp}.  
Moreover, the maximal value of $\Delta$ that determines whether weak-scale
SUSY is a fine-tuned model (should it be $\Delta\sim 10$?
100? 1000?) is ultimately subjective. 
Thus, it is premature to conclude that weak-scale SUSY is on
the verge of exclusion.  It might be possible to sharpen
the upper bounds on superpartner masses based on naturalness
arguments, which ultimately will either confirm or refute the weak-scale
SUSY hypothesis\cite{Baer:2015rja}{}.  Of course,
if evidence for supersymmetric phenomena in the multi-TeV regime were
to be established at a future collider facility (with an energy reach
beyond the LHC\cite{susy1:FCChh}{}), it would be viewed as a
spectacularly successful explanation of the large gauge hierarchy
between the (multi-)TeV scale and Planck scale.  In this case, the
remaining little hierarchy, characterized by the somewhat large value
of the fine-tuning parameter $\Delta$ discussed above, would be
regarded as a less pressing issue.

\subsection{Indirect constraints on supersymmetric models}\label{susy1:sec:constr-virt-exch}

While direct empirical searches for supersymmetric particles provide various
limits on their properties, indirect constraints can depend more sensitively on details of the
whole model. The cold dark matter relic density inferred from
cosmological fits to observational data is one
such example of an indirect constraint. In supersymmetric models
where the LSP is stable (and thus is a dark matter candidate), its thermally-produced relic density
depends upon the scattering of various supersymmetric particles into dark matter
particles and SM particles. The resulting relic density can depend sensitively
on the masses of the non-LSP supersymmetric particles as well as on the mass
of the LSP.
In a typical model, an appreciable region of the parameter
space is ruled out because it yields 
an overabundance of dark matter (see for example~\Ref{GAMBIT:2017zdo} for a
fit to a seven parameter version of the pMSSM).
However, subsequent tweaks to the supersymmetric model that yield an unstable LSP, such as the introduction of
R-parity violating effects, can mean that the
relic density no longer constrains the parameter space.

There are a number of indirect constraints based on low-energy measurements that are sensitive to the effects of new physics via
supersymmetric loop effects.
For example, the virtual exchange of supersymmetric particles
can contribute to the muon anomalous magnetic moment,
$a_\mu\equiv\frac{1}{2}(g-2)_\mu$, as reviewed in~\Ref{Athron:2025ets}. 
A current SM prediction for $a_\mu$~\cite{Aliberti:2025beg} uses conventional perturbative calculations,  employing lattice QCD calculations for the hadronic vacuum polarisation contribution as well as dispersive methods \emph{and} lattice QCD estimates of the hadronic light-by-light contribution. The resulting prediction displays no significant tension with the experimentally observed value~\cite{Muong-2:2025xyk}. This is in contrast to previous estimates that relied on dispersive methods for estimating the hadronic vacuum polarization contribution and found a significant deviation between the measured value and the SM prediction (see \eg \Ref{Aoyama:2020ynm}).

The precision of the measured value of $a_\mu$ is not sensitive to the 
experimental error associated with the measured value of the fine structure constant, $\alpha$.  In contrast, the comparison of the 
SM prediction with the experimental measurement of the anomalous magnetic moment of the electron, $a_e$, depends critically on the value of $\alpha$.
Using the experimentally determined value of $\alpha$ given in \Ref{Hanneke:2008tm}
yields a SM prediction for $a_e$ that is 2.4$\sigma$ above its measured value~\cite{Parker:2018vye}.
However,
this previous determination of $\alpha$ is in tension at the 5$\sigma$
level with a more recent measurement of the fine structure
constant~\cite{Morel:2020dww}.  The latter yields 
a SM prediction
for $a_e$ that is 1.6$\sigma$ below its measured value~\cite{Morel:2020dww}.
Measurements of the fine
structure constant, $a_e$ and $a_\mu$ jointly
constrain the pMSSM parameter
space~\cite{Dutta:2018fge,*Endo:2019bcj,*Badziak:2019gaf,*Li:2021koa} due to
shifts originating from supersymmetric loop effects.

Flavor transitions in radiative, leptonic and semi-leptonic
\index{flavor anomalies}%
$b$ quark decays~\cite{Albrecht:2021tul} provide a fertile ground for physics
beyond the SM\@. For example,
the rare inclusive decay
$b\to s\gamma$ is a sensitive probe of the virtual
effects of new physics beyond the SM\@.  The experimental
measurements of $B\to X_s+\gamma$\cite{Lees:2012ym,*Lees:2012ufa,*Belle:2014nmp}
are in agreement with the theoretical
SM predictions of \Ref{Misiak:2015xwa,*Czakon:2015exa,*Misiak:2020vlo}.
Since supersymmetric loop corrections can contribute an
observable shift from the SM predictions,
the absence of any significant deviation places useful constraints on
the MSSM parameter space\cite{Baer:1996kv,*Ciuchini:2002uv,*Hurth:2003vb,*Mahmoudi:2007gd,*Olive:2008vv}.

The rare decays $B_s\to\mu^+\mu^-$ and $B_d\to \mu^+\mu^-$ are especially sensitive to
supersymmetric loop effects, with some loop contributions scaling
as $\tan^6\beta$ when $\tan\beta\gg 1$\cite{Choudhury:1998ze,*Babu:1999hn,*Isidori:2001fv,*Isidori:2002qe}.  
Since there are no significant deviations observed between the measurements~ of these rare decay modes~\cite{CMS:2022mgd,*Aaboud:2018mst,*LHCb:2021awg} and the predicted SM rates~\cite{Bobeth:2013uxa}, constraints on the parameter space of the MSSM can be derived~\cite{Choudhury:1998ze,*Babu:1999hn,*Isidori:2001fv,*Isidori:2002qe}.

Several tensions exist between SM predictions and measurements of some other
experimental 
observables that probe $b\rightarrow s \mu^+\mu^-$ transitions, although
the level of tension depends upon the theoretical treatment of the SM
analysis. 
In a certain angular distribution parameter (denoted by $P_5^\prime$)
extracted from $B^0 \rightarrow K^{\ast0} \mu^+ \mu^-$
decays, the tension is around the 4$\sigma$ level~\cite{Aaij:2020nrf}. An even
larger discrepancy is observed in
a combination of angular distributions and rates derived from 
$B^\pm \rightarrow K^\pm \mu^+ \mu^-$ and $B^0 \rightarrow
K^0 \mu^+ \mu^-$~\cite{Gubernari:2022hxn}.  
Moreover, there is a 3.6$\sigma$ deviation in the branching ratio of
$B_s \rightarrow \phi \mu^+\mu^-$ for di-muon invariant mass squared
values between 1.1 GeV$^2$ and 6.0 GeV$^2$~\cite{LHCb:2021zwz}.
Finally, a recent measurement of $B^+ \rightarrow K^+\nu \bar \nu$ by the
Belle II Collaboration~\cite{Belle-II:2023esi}
obtained a branching fraction that is roughly four times larger than
the SM prediction~\cite{Parrott:2022zte}, corresponding to a
 2.7$\sigma$ deviation from the SM expectation.
However, it is unlikely that this result can be attributed to new
contributions from the MSSM~\cite{Bertolini:1990if,*Aliev:1993bc}, as
the latter are expected to be negligible in comparison with the SM.  

The decays $B^\pm\to\tau^\pm\nu_\tau$ and $\overline{B}\to
D^{(*)}\tau^-\overline\nu_\tau$
are noteworthy, since in models with extended Higgs sectors such as the MSSM, these processes
possess tree-level charged Higgs exchange contributions that can
compete with the dominant $W$-exchange.  As shown in Ref.~\cite{PDGleptondecay},
experimental measurements of 
$B^\pm\to\tau^\pm\nu_\tau$
are currently consistent with SM expectations\cite{Bona:2009cj}. 
\index{Collaborations!BaBar}%
The BaBar Collaboration measured values of the rates
for $\overline{B}\to D\tau^-\overline\nu_\tau$ and $\overline{B}\to
D^*\tau^-\overline\nu_\tau$\cite{Lees:2012xj,*Lees:2013uzd}
which exhibited a combined
3.4$\sigma$ discrepancy from the SM predictions, which was
also not compatible with the Type-II Higgs Yukawa couplings employed
by the MSSM\@.  
\index{Collaborations!LHCb}%
\index{Collaborations!Belle}%
Subsequent measurements by the LHCb and Belle
Collaborations
were compatible with the BaBar measurements although they
displayed less deviation
from the SM expectations.
The combined
difference between the measured values of the
$\overline{B}\to D\tau^-\overline\nu_\tau$ and $\overline{B}\to
 D^*\tau^-\overline\nu_\tau$ decay rates relative to the corresponding
 SM values has a significance of $3.1\sigma$\cite{HeavyFlavorAveragingGroupHFLAV:2024ctg}.

In summary, although there are a few hints of possible deviations from
the SM in $B$ decays, none of the discrepancies by
themselves are significant enough to conclusively imply the existence
of new physics beyond the SM\@.
Moreover, the absence of evidence for sizable deviations in other $B$-physics 
observables from their SM predictions can
place useful constraints on the MSSM parameter space\cite{Carena:2008ue,Buchmueller:2013rsa,Mahmoudi:2012un,*Arbey:2017eos,*Eberl:2021ulg}.

\subsection{Standard Model Effective Field Theory and SUSY \label{susy1:sec:smeft}}

The SM effective field theory (SMEFT)~\cite{Isidori:2023pyp} 
\index{SMEFT, SM effective field theory}%
\index{SM effective field theory, SMEFT}%
encodes the effects of fields beyond those of the SM under the assumption that the characteristic mass scale $\Lambda$ associated with the new fields is much higher than the highest energy scale $E$ of some set of experimentally probed observables. The total Lagrangian density is written
\begin{equation}
{\mathcal L} = {\mathcal L}_4 + \sum_{d=5}^\infty \sum_i \frac{C_i}{\Lambda^{d-4}} {\mathcal O}_i^{(d)}, \label{eq:susy1:SMEFT}
\end{equation}
where ${\mathcal L}_4$ is the usual renormalizable Lagrangian density of the SM, ${\mathcal O}_i^{(d)}$ is an operator of mass dimension $d$ composed of a product of SM fields (and their spacetime derivatives), and the $C_i$ are dimensionless Wilson coefficients.  The index $i$ labels independent operators (for an example of an operator basis up to $d=6$, see \Ref{Grzadkowski:2010es}). The summed term of \Eq{eq:susy1:SMEFT} arises from integrating out fields with masses of order $\Lambda$ (or larger) from the theory. The effects of successively higher-dimension operators on an observable are suppressed by a factor $(E/\Lambda)^{d-4}$.
Thus, in a set of SMEFT operators that contribute to an observable, the dominant effects come from those operators with smallest dimension $d$.
Experimental measurements can then be interpreted as bounds on the $C_i$, provided that no on-shell beyond-the-SM resonances affect the measurements. 
After electroweak symmetry breaking, the $d=5$ operators describe neutrino mass terms and so $d=6$ SMEFT operators are generically the most relevant for collider physics.  

Measurements involving SM particles 
are often interpreted as constraints on the Wilson coefficients of $d=6$ operators. 
The MSSM has been matched to the $d=6$ SMEFT Lagrangian at the one-loop level in \Ref{Kraml:2025fpv}.
In principle, the bounds on the $C_i$ obtained from experimental studies can be translated into constraints on the MSSM parameter space.

\section{Massive neutrinos in weak-scale SUSY}\label{susy1:sec:massive-neutral}

In the minimal version of the SM and its supersymmetric extension,
there are no right-handed neutrinos, and Majorana mass terms for
the left-handed neutrinos are absent.  However, given
the overwhelming evidence for neutrino masses and
mixing (see Refs.~\cite{PDGnu,deSalas:2020pgw,*Esteban:2020cvm}), any viable model of the fundamental particles must
provide a mechanism for generating neutrino masses\cite{Zuber:1998xe,*King:2015aea,*King:2017guk}{}.
In extended supersymmetric models, various mechanisms exist for producing
massive neutrinos\cite{susy1:susynureview,*Hirsch:2004he}{}.  Although one can
devise models for generating massive Dirac neutrinos\cite{Borzumati:2000mc}{},
the most common approaches for incorporating neutrino masses are based on
$L$-violating supersymmetric extensions of the MSSM, which generate
massive Majorana neutrinos.  Two classes of $L$-violating supersymmetric
models will now be considered.

\subsection{The supersymmetric seesaw}\label{susy1:sec:susy-seesaw}

Neutrino masses can be incorporated
into the SM
by introducing SU(3)$\times$SU(2)$\times$U(1)
singlet right-handed neutrinos ($\nu_R$)
whose mass parameters are very large,
typically near the grand unification scale.
In addition, one must also include a standard Yukawa
couplings between the lepton doublets, the Higgs doublet, and $\nu_R$.
The Higgs VEV  then induces an off-diagonal
$\nu_L$--$\nu_R$ mass on the order of the electroweak scale.  Diagonalizing
the neutrino mass matrix (in the three-generation model)
yields three superheavy neutrino states, and three
very light neutrino states that are identified with the light neutrinos
observed in nature.  This is the seesaw mechanism\cite{Minkowski:1977sc,*susy1:seesaw2,*Yanagida:1980xy,*Mohapatra:1979ia,*Mohapatra:1980yp}.

It is straightforward to construct a
supersymmetric generalization of the seesaw model
of neutrino masses\cite{Hisano:1995nq,*Hisano:1995cp,*Casas:2001sr,*Ellis:2002fe,*Masiero:2004js,*Arganda:2004bz,*Joaquim:2006uz,*Ellis:2007wz,Grossman:1997is,*Dedes:2007ef}
by promoting the right-handed neutrino field to
a superfield $\widehat N^c=(\widetilde\nu_R\,;\,\nu_R)$.
Integrating out the heavy right-handed neutrino supermultiplet yields a new term in the
superpotential [cf.~\Eq{susy1:eq:Wsup}] of the form
\begin{equation}\label{susy1:eq:Wseesaw}
W_{\rm seesaw}= \frac{f}{M_R} (\widehat H_U\widehat L) (\widehat H_U\widehat L)\,,
\end{equation}
where $M_R$ is the mass scale of the right-handed neutrino sector and $f$ is a dimensionless
constant.
Note that lepton
number is broken by two units by \Eq{susy1:eq:Wseesaw}, which implies that
R-parity invariance is preserved.
The supersymmetric analogue of the Majorana neutrino mass term
in the sneutrino sector leads to
sneutrino--antisneutrino mixing
phenomena\cite{Grossman:1997is,*Dedes:2007ef,Hirsch:1997vz,*Hall:1997ah,*Choi:2001fka,*Honkavaara:2005fn}.
 In addition, new Higgs-slepton interaction terms can probe the structure of
 the supersymmetric seesaw model~\cite{Liu:2023ydb}.
The right-handed sneutrino that resides in $\widehat N^c$ also provides an
intriguing dark matter candidate~\cite{Faber:2019mti}. 

The SUSY Les Houches Accord\cite{Allanach:2008qq,Skands:2003cj}{}, mentioned at
the end of the introduction to
\Sec{susy1:sec:sp-spectrum}, has been extended to the
supersymmetric seesaw (and other extensions of the MSSM) in \Ref{Basso:2012ew}.

\subsection{R-parity-violating SUSY}\label{susy1:sec:R-parity-viol-susy}
\index{RPV, R-parity violation}%
\index{R-parity violation, RPV}%
It is possible to incorporate massive neutrinos in renormalizable
supersymmetric models while
retaining the minimal particle content of the MSSM by relaxing  the assumption of
R-parity invariance.
The most general
R-parity-violating model involving the MSSM spectrum introduces
many new parameters to both the SUSY-conserving and the
SUSY-breaking sectors\cite{Chemtob:2004xr,*Barbier:2004ez,Allanach:2008qq}{}.  Each new interaction term
violates either $B$ or $L$ conservation.   For example, starting
from the MSSM superpotential given in \Eq{susy1:eq:Wsup} [suitably generalized
to three generations of quarks, leptons and their superpartners],
consider the effect of adding the following new
terms:
\begin{equation}\label{susy1:eq:rpv}
\begin{aligned}
W_{\rm RPV}=&
(\lambda_L)_{pmn} \widehat L_p \widehat L_m \widehat E^c_n
+ (\lambda_L^\prime)_{pmn}\widehat L_p \widehat Q_m\widehat D^c_n 
\bookcr{}{&}
+(\lambda_B)_{pmn}\widehat U^c_p \widehat D^c_m \widehat D^c_n
+(\mu_L)_p \widehat H_u\widehat L_p\,,
\end{aligned}
\end{equation}
where $p$, $m$, and $n$ are generation indices, and
gauge group indices are suppressed.
\Eq{susy1:eq:rpv} yields new scalar-fermion Yukawa couplings consisting of
all possible combinations involving two SM fermions
and one scalar superpartner.
Note that the term in \Eq{susy1:eq:rpv}
proportional to $\lambda_B$ violates $B$, while the other three terms
violate $L$.
The $L$-violating term in \Eq{susy1:eq:rpv} proportional to $\mu_L$ is the RPV analog of the
$\mu \widehat H_u\widehat H_d$ term of the MSSM superpotential,
in which the $Y=-1$ Higgs/higgsino supermultiplet $\widehat H_d$ is replaced
by the slepton/lepton supermultiplet $\widehat L_p$.

Phenomenological constraints derived from data on
various low-energy $B$- and $L$-violating processes can be used to
establish limits on each of the coefficients
$(\lambda_L)_{pmn}$, $(\lambda_L^\prime)_{pmn}$, and
$(\lambda_B)_{pmn}$ taken one at a time\cite{Chemtob:2004xr,*Barbier:2004ez,susy1:dreiner}{}.
If more than one coefficient is simultaneously non-zero, then the limits
are in general more complicated\cite{Allanach:1999ic}{}.
All possible RPV terms
cannot be simultaneously present and unsuppressed; otherwise
the proton decay rate would be many orders of magnitude larger than
the present experimental bound.   One way to avoid proton decay is to
impose $B$ or $L$~invariance (either one alone would suffice).  Otherwise,
one must accept the requirement that
certain RPV coefficients must be extremely suppressed.

One particularly interesting class of RPV models is one in which $B$ is
conserved, but $L$ is violated.  It is possible to enforce baryon number
conservation (and the stability of the proton),
while allowing for lepton-number-violating interactions
by imposing a discrete $\mathbb{Z}_3$ baryon triality
symmetry on the low-energy
theory\cite{Ibanez:1991pr,*Ibanez:1992ji}{}, in place of the standard $\mathbb{Z}_2$
R-parity.
Since the distinction between the Higgs and matter
supermultiplets is lost in RPV models where $L$ is violated,
the mixing of sleptons
and Higgs bosons, the mixing of neutrinos and neutralinos, and the
mixing of charged leptons and charginos are now possible,
leading to more complicated
mass matrices and mass eigenstates than in the MSSM\@.
The treatment of neutrino masses and mixing in this framework
can be found, {\it e.g.}, in \Ref{Dedes:2006ni,*Allanach:2007qc,*Dreiner:2011ft}.

Alternatively, one can consider imposing a lepton parity such that all
lepton superfields are odd\cite{Ibanez:1991pr,Dreiner:2005rd}{}.
In this case, only the $B$-violating term
in \Eq{susy1:eq:rpv} survives, and $L$ is conserved.  Models of this type have
been considered in \Ref{Tamvakis:1996np,*Eyal:1999gq,*Florez:2013mxa}.  Since $L$ is conserved in these models, the mixing
of the lepton and Higgs superfields is forbidden. Moreover, neutrino masses
(and mixing) are not generated if lepton parity is an exact symmetry.
However, one
expects that lepton parity cannot be exact due to
quantum gravity effects.
Remarkably, the standard $\mathbb{Z}_2$ R-parity and the $\mathbb{Z}_3$
baryon triality are stable with respect to quantum gravity effects,
as they can be identified as
residual discrete symmetries that arise from spontaneously broken
non-anomalous gauge symmetries\cite{Ibanez:1991pr}.
The symmetries employed above to either remove or suppress R-parity
violating operators were flavor independent. In contrast, there exist a
number of motivated scenarios based on flavor symmetries that can also yield
the suppression as required by the experimental data ({\it e.g.}, see \Ref{Csaki:2011ge}).

The supersymmetric phenomenology of the RPV models exhibits features
that are distinct from that of the MSSM\cite{Chemtob:2004xr,*Barbier:2004ez}{}.  The
LSP is no longer stable, which implies that not all supersymmetric
decay chains must yield missing-energy events at colliders.  A comprehensive
examination of the phenomenology of the MSSM extended by a single
R-parity violating coupling at the unification scale and its
implications for LHC searches has been given in
\Ref{Dercks:2017lfq}.
As an example, the sparticle mass bounds obtained
in searches for R-parity-conserving SUSY can be considerably
relaxed in certain RPV models due to the absence of large missing transverse
momentum signatures\cite{Allanach:2012vj,*Asano:2012gj,*Chamoun:2014eda}{}.  This can alleviate some of the
tension with naturalness (discussed in \Sec{susy1:sec:natural-constr}).

Nevertheless, the loss of the missing-energy signature is often
compensated by other striking signals (which depend on which
R-parity-violating terms are dominant).  For example,
supersymmetric particles in RPV models can be singly produced (in
contrast to R-parity-conserving models where supersymmetric particles
must be produced in pairs).  The phenomenology of pair-produced
supersymmetric particles is also modified in RPV models
due to new decay chains not present in R-parity-conserving
SUSY models\cite{Chemtob:2004xr,*Barbier:2004ez}.

In RPV models with lepton number violation (these include
weak-scale SUSY models with baryon triality mentioned above),
both $\Delta L\!=\! 1$ and
$\Delta L\!=\! 2$ phenomena are allowed, leading to neutrino masses
and mixing\cite{Romao:1999bu,*Grossman:2003gq}{}, neutrinoless double-beta
decay\cite{Mohapatra:1986su,*Babu:1995vh,*Hirsch:1995zi,*Hirsch:1995ek}, sneutrino-antisneutrino
mixing\cite{Grossman:1998py}{}, and resonant \hbox{$s$-channel}
production of sneutrinos in $e^+e^-$
collisions\cite{Dimopoulos:1988jw,*Kalinowski:1997bc,*Erler:1996ww} and in
charged sleptons in $p\bar p$ and $pp$
collisions\cite{Dreiner:2000vf}, respectively.

\section{Extensions beyond the MSSM}
\label{susy1:sec:extensions}

In addition to possible extensions of the MSSM to incorporate massive neutrinos discussed in \Sec{susy1:sec:massive-neutral}, there are numerous reasons for considering more general
extensions of the MSSM\cite{susy1:Moretti}.
Possible extensions include an
enlarged electroweak gauge group beyond
SU(2)$\times$U(1)\cite{Hewett:1988xc,Choi:2006fz,Allanach:2021yjy}{}, 
the addition of new Higgs supermultiplets beyond the doublets and singlets of
the MSSM/NMSSM\cite{Delgado:2012sm}{},
and/or the addition of new (possibly exotic) matter 
supermultiplets\cite{King:2005my,*King:2005jy,Martin:2012dg,Kawase:2011az,*Escudero:2005hk,*FileviezPerez:2012iw,*Dutta:2018yos,*Altmannshofer:2021hfu}
such as vector-like fermions and their superpartners or adjoint chiral superfields\cite{Nelson:2002ca,Benakli:2011kz,*Benakli:2011vb}.  

In this final Section, we shall briefly focus on extensions of the MSSM that have been proposed to solve specific
theoretical problems associated with the MSSM.

\subsection{The origin of the $\mu$ parameter}
\label{muproblem}

In the MSSM, the parameter $\mu$ that appears in \Eq{susy1:eq:Wsup} is 
a SUSY-{\it preserving}\/ parameter.  However, its magnitude
must be of order the effective SUSY-breaking scale
of the MSSM to yield a
consistent supersymmetric phenomenology\cite{Kim:1983dt}.
Any natural solution to the so-called $\mu$-problem must incorporate
a symmetry that enforces $\mu=0$ and a small symmetry-breaking
parameter that generates a value of $\mu$ that is not parametrically
larger than the effective SUSY-breaking
scale\cite{Kim:1994eu}.  A number of proposed mechanisms in the
literature ({\it e.g.}, see \Ref{Kim:1983dt,Kim:1994eu,Giudice:1988yz,*Casas:1992mk,*Bae:2019dgg}) provide
concrete examples of a natural solution to the $\mu$-problem of the MSSM\@.

In extensions of the MSSM, other compelling solutions to the
$\mu$-problem are possible.
For example, one can replace $\mu$ by the
VEV of a new SU(3)$\times$SU(2)$\times$U(1)
singlet scalar field, as noted below \Eq{susy1:eq:WsupN}.  This is the NMSSM, which yields phenomena
that were briefly discussed in
Sections~\ref{susy1:sec:sp-spectrum}--\ref{susy1:sec:exp-MSSM}.
The NMSSM superpotential consists only of trilinear terms whose coefficients
are dimensionless.
One can also extend 
the NMSSM further to the so-called USSM\cite{Choi:2006fz} by
adding a new broken U(1) gauge symmetry\cite{Cvetic:1997ky},
under which the singlet field is charged.

Alternatively, one can consider a generalized
\index{GNMSSM, generalized NMSSM}%
\index{generalized NMSSM, GNMSSM}%
version of the NMSSM (called the GNMSSM in \Ref{Ross:2011xv}),
where all possible renormalizable terms in the superpotential are allowed,
which yield new supersymmetric mass terms (analogous to the $\mu$
term of the MSSM).  A discussion of the parameters of the GNMSSM can
be found in \Ref{Allanach:2008qq}.
Although the GNMSSM does not solve the
$\mu$-problem, it does exhibit regions of parameter space in which
the degree of fine-tuning is relaxed, as mentioned in \Sec{susy1:sec:natural-constr}.

The generation of the $\mu$-term may be connected with the solution to the
strong CP problem\cite{Peccei:2006as}{}.  \index{mu@$\mu$-term}
Models of this type, which include
new gauge singlet fields that are charged under
the Peccei-Quinn (PQ) symmetry\cite{Peccei:1977hh,*Peccei:1977ur}{}, were first proposed in \Ref{Kim:1983dt}.
The breaking of the PQ symmetry is thus intimately tied to SUSY breaking,
while naturally yielding a value of $\mu$ that is of order the electroweak symmetry breaking 
scale\cite{Murayama:1992dj,*Gherghetta:1995jx,*Bae:2014yta,*Baer:2018avn}.

\subsection{Dirac gauginos}
\label{DiracGauginos}

The MSSM and its extensions considered so far in this review contain
self-conjugate fermions---the Majorana gluinos and neutralinos.
The possibility of Dirac gluinos and neutralinos arises if 
the R-parity of the supersymmetric model is promoted to a continuous U(1)$_R$ symmetry.  In the MSSM, this could be (partially) achieved by setting the supersymmetry-conserving higgsino mass parameter $\mu$ and the supersymmetry-breaking gaugino Majorana mass parameters $M_i$ ($i=1,2,3$) and $A$-parameters to zero.
The resulting spectrum features a massless gluino and a pair of massive Dirac neutralinos whose masses are generated at the tree-level and at the one-loop level, respectively~\cite{Hall:1990hq,*Randall:1992cq}.

The mass spectrum of gluinos/neutralinos quoted above
is not phenomenologically viable.
Nevertheless, one can devise more realistic models 
by expanding the MSSM to include
additional chiral superfields in the
adjoint representation (\eg, a color octet chiral superfield), while still respecting the U(1)$_R$ symmetry.  The spin-1/2 components of the adjoint chiral
superfields can pair up with the gauginos to form Dirac gauginos\cite{Fayet:1978qc,Hall:1990hq,*Randall:1992cq,Nelson:2002ca,Benakli:2011kz,*Benakli:2011vb}{}.  Such states appear
\index{Dirac gauginos}%
in models of so-called supersoft SUSY
breaking\cite{Fox:2002bu}{} and in some generalized GMSB models\cite{Nelson:2002ca,Benakli:2008pg,*Benakli:2010gi}.
Moreover, by introducing additional Higgs supermultiplets while maintaining the U(1)$_R$ symmetry, the spin-1/2 components of these new superfields can pair up with the neutral higgsino states of the MSSM to form Dirac neutralinos~\cite{Kribs:2007ac}.
Various scenarios, often referred to as constrained minimal Dirac gaugino supersymmetric models, have been proposed and analyzed in Refs.~\cite{Kribs:2007ac,Benakli:2014cia,*Goodsell:2020lpx,*Benakli:2022gjn}.

The phenomenology of models with Dirac gauginos can be quite distinct from that of the MSSM, as noted in \Ref{Chalons:2018gez}.
Moreover, such models can lead to significantly
relaxed flavor constraints~\cite{Kribs:2007ac}.  
In particular, the higher mass scale of the gauginos and their Dirac nature lead to suppressed colored sparticle production at the LHC~\cite{Kribs:2012gx}.
The implications of models of Dirac gauginos
for the properties of the observed Higgs boson and  possibilities for improved naturalness
are addressed in \Ref{Benakli:2012cy,*Bertuzzo:2014bwa,*Nakano:2015gws}.

\bigskip
\section*{Acknowledgments}
\medskip
We would like to thank Wolfgang Altmannshofer, Peter Athron, Howard Baer,
Herbi Dreiner, Sabine Kraml, Pietro Slavich, and Dominik St\"{o}ckinger for a number of useful suggestions.  The work of B.A. is partially supported by the STFC consolidated grant ST/X000664/1.
B.A. is grateful for the hospitality and support during his visit to the
Santa Cruz Institute for Particle Physics at the University of California,
Santa Cruz, where some of the work on the 2025 update of this review was carried
out.   The work of H.E.H. is supported in part by the U.S. Department of Energy Grant No.~\uppercase{DE-SC}0010107.
H.E.H. is also grateful for the hospitality and support during his visits to the Department of Applied Mathematics and Theoretical Physics at the University of
Cambridge, where some of the work on the 2023 review and its subsequent update were carried
out. 

\bigskip\bigskip
\phantomsection
\centerline{\bf \large References}

\addcontentsline{toc}{section}{References}
\bibliographystyle{pdg}
\bibliography{susy1-pdg.bib}

\newpage

\phantomsection
\centerline{\bf \large Index} 
\bigskip\bigskip

\titleformat{\chapter}[display]{}{}{}{\centering \normalsize \selectfont\mbox}
\titlespacing*{\chapter}{0pt}{*-5.7}{0pt}
\addcontentsline{toc}{section}{Index}
\renewcommand{\indexname}{}

\renewcommand{\twocolumn}[1][]{
     \twocolumngrid
     #1
}

\printindex

\end{document}